\documentclass[preprintnumbers,amssymb,prd, notitlepage,nofootinbib,twocolumn, superscriptaddress]{revtex4-2}

\newcommand{\apjl}{Astrophys. J. Lett.}
\newcommand{\aap}{Astron. Astrophys.}
\newcommand{\apjs}{Astrophys. J. Suppl. Ser.}

\newcommand{\mnras}{Mon. Not. R. Astron. Soc.}

\newcommand{\jcap}{J. Cosmol. Astropart. Phys.}

\usepackage{graphicx}
\usepackage{amsmath,amsbsy,pifont, amssymb}
\usepackage{multirow}
\usepackage{soul, xcolor}
\newcommand{\comment}[1]{{}}
\newcommand{\pa}{\partial}
\newcommand{\en}{\nonumber\\}
\newcommand{\lcdm}{\ensuremath{\Lambda\mathrm{CDM}}}

 \newcommand{\cmark}{\ding{51}}
 \newcommand{\xmark}{\ding{55}}
 \def\be{\begin{equation}}
\def\ee{\end{equation}}
\def\ba{\begin{eqnarray}}
\def\ea{\end{eqnarray}}
 
\begin{document}

\title{Scaling Transformations and the Origins of Light Relics Constraints from Cosmic Microwave Background Observations}

\author{Fei Ge}
\affiliation{Department of Physics and Astronomy, University of California, Davis, California 95616, USA}
\author{Francis-Yan Cyr-Racine}
\affiliation{Department of Physics and Astronomy, University of New Mexico, Albuquerque, New Mexico 87106, USA}
\author{Lloyd Knox }
\affiliation{Department of Physics and Astronomy, University of California, Davis, California 95616, USA}

\email{fge@ucdavis.edu}

\date{\today}

\begin{abstract}
We use here a family of scaling transformations, that scale key rates in the evolution equations, to analytically understand constraints on light relics from cosmic microwave background (CMB) maps, given cosmological models of varying degrees of complexity. We describe the causes of physical effects that are fundamentally important to the constraining power of the data, with a focus on the two that are most novel. We use as a reference model a cosmological model that admits a scaling transformation that increases light relic energy density while avoiding all of these causes. Constraints on light relics in a given model can then be understood as due to the differences between the given model and the reference model, as long as the additional light relics only interact gravitationally with the Standard Model components. This understanding supports the development of cosmological models that can evade light relics constraints from CMB maps.

\end{abstract}


\maketitle

\section{Introduction} \label{sec:intro}

Extensions of the Standard Model of particle physics often contain light degrees of freedom (see e.g.~Refs.~\cite{Peccei:1977hh,Weinberg:1977ma,Wilczek:1977pj,Feng:1997tn,Chacko:2005pe,Dvali:2007hz,Dvali:2009ne,Arkani-Hamed:2016rle,Chacko:2018vss}), in addition to those of the photon and standard model neutrinos, that would be thermally produced in the early universe. Constraints on such scenarios have come for decades from observations of the abundances of light elements, helium \cite{Izotov:2014fga,2018MNRAS.478.5301F,Aver:2015iza,Hsyu:2020uqb,aver2021improving} and deuterium \cite{Cooke:2016rky,Riemer-Sorensen:2017pey,Cooke:2017cwo,Yeh:2020mgl} in particular.
More recently, observations of the cosmic microwave background \cite{Planck:2018vyg,SPT-3G:2021wgf,aiola:2020} have also led to constraints that likewise have an important influence on particle physics model building. For a recent review of current constraints on light relic abundances and their associated physical models see Ref.~\cite{Dvorkin:2022}. Precision measurement of the energy density in light relics is a major science goal of next-generation cosmic microwave background (CMB) projects such as Simons Observatory \cite{SimonsForecast} and CMB-S4 \cite{Abazajian:2022}. 

Despite intense interest in constraining light relics, and decades of study of such constraints, the existing literature is incomplete in its description of how these constraints arise from CMB observations. 
Useful explanations do exist in the literature (see e.g.~Refs.~\cite{Hu:1996,Bashinsky:2003tk,Hou:2011ec,Baumann:2015rya,Follin:2015hya}), but they leave us unable to answer certain questions. The desire to answer these questions is what motivated our work and has led us to the more complete understanding of the origins of light-relics light-relic constraints that we present here.

Our analysis has focused on all the dimensionful quantities in the relevant Einstein-Boltzmann equations governing evolution of the metric, matter, and radiation. These are: the gravitational free-fall rates, $\sqrt{G\rho_i(z)}$ for each component $i$ , the photon (Thomson) scattering rate $\sigma_{\rm T}n_{\rm e}(z)$, hydrogen recombination rates and the Fourier mode wave number $k$. For scale-invariant initial conditions, a uniform scaling of all the rates and $k$, leaves dimensionless observables invariant \cite[][hereafter CGK]{Cyr-Racine:2022}, a conclusion that can be reached by dimensional analysis alone. Invariance under a power-law primordial power spectrum can be achieved by adding an appropriate scaling of its amplitude \cite{Zahn_2003,Cyr-Racine:2022}. In a given model space, the energy density in light relics can be increased as part of a uniform scaling up of a number of these rates. If they can all be scaled then there will be no constraints on light relic densities from dimensionless observables. In general, only some of the rates can be uniformly scaled and those model spaces that admit only less-comprehensive scaling transformations will have tighter constraints on light relics. 

In CGK we also introduced a physical model in which a scaling transformation could be implemented that, while not leading to an exact symmetry, had very small symmetry-breaking effects.  In this paper we expand upon our previous description of the scaling transformation and use it to improve our understanding of constraints on light relics. We use the physical model as a reference model that supports our analytic understanding of results in other model spaces that also have additional light relics. 

Previous explanations \cite{Hu:1996,Hou:2011ec}, although they did not use this exact language, took advantage of a scaling transformation, $H(z) \rightarrow \lambda H(z)$, achievable in standard model extensions with additional light relics, under which there is an approximate symmetry of angular scales strongly linked with observables: that of the sound horizon, $\theta_{\rm s}$ and of the Hubble length scale at matter-radiation equality, $\theta_{\rm EQ}$ (both projected from the last-scattering surface). With this scaling transformation as a starting point, the effects that break the symmetry, rendering it only approximate, can be understood as leading to our ability to constrain additional energy density in light relics. Photon diffusion is such an effect; the angular scale of the photon diffusion distance at last scattering is not invariant under a uniform scaling of the Hubble rate \cite{Zaldarriaga:1995gi}. Photon diffusion breaks the symmetry, giving us sensitivity to $H(z)$, as long as there are no other parameters to vary in the model that could significantly impact photon diffusion.

This damping scale effect is one of several effects leading to constraints on light relics. We list here four causes of such effects, all arising from nonuniform changes to the rates we listed above. Three of these have already been articulated in the literature and one we articulate here for the first time. These are
\begin{enumerate}
\item changes to the photon scattering rate relative to the Hubble rate\footnote{More precisely we mean here all changes to the photon scattering rate relative to the Hubble rate at fixed ionization fraction $x_{\rm e}(z)$.}: leads to damping and polarization generation effects \cite{Hu:1996,Zahn_2003,Martins:2010gu,Hou:2011ec}, 
\item changes to the fraction of radiation density that is free-streaming particles prior to recombination: changes acoustic oscillation phase and amplitude \cite{Bashinsky:2003tk,Baumann:2015rya,Follin:2015hya}, 
\item changes to the fraction of matter that is pressure-supported: leads to gravitational potential differences that impact photon perturbation evolution (effects we describe here for the first time), and
\item changes to recombination rates relative to the Hubble rate: impacts the ionization history $x_{\rm e}(z)$ with ensuing observational consequences \cite{Zahn_2003}. 
\end{enumerate} 
All of these can be understood as arising from departures from the scaling transformation presented in CGK. The causes are in a decreasing order of the magnitude of symmetry-breaking impact they can have. 

This work was stimulated by our desire to understand the origin of constraints in a model space that had degeneracy directions that could eliminate all effects from causes 1 and 2. Such constraints were presented in Refs.~\cite{Baumann:2015rya,Brust:2017nmv,Blinov:2020hmc,Brinckmann:2020bcn}. There one can see that even if the photon scattering rate to Hubble rate ratio can be held constant via a change in the primordial helium abundance, and if the free-streaming fraction can be held fixed, the constraints on light relics, while loosened, still exist. 

The scaling transformation presented in CGK, by allowing for the elimination of causes 1, 2, and 3, provides us with a convenient means of understanding the constraints presented in  Refs.~\cite{Baumann:2015rya,Brust:2017nmv,Blinov:2020hmc,Brinckmann:2020bcn}. These constraints can then be understood as arising from differences between the models considered in these works and the reference model exploiting the scaling transformation presented in CGK. We use this approach to articulate what we have enumerated as cause 3 and its associated effects.  

Scaling transformations as a means to understand constraints on new physics have been used elsewhere in the literature, most notably by Zahn and Zaldarriaga \cite[][hereafter ZZ]{Zahn_2003}. They were not trying to understand constraints on light relics in particular, but more generally the sensitivity to the expansion rate through the epoch of recombination. To that end they introduced a scaling of Newton's constant $G \rightarrow \lambda^2 G$, and showed that CMB temperature and polarization anisotropy measurements at small scales, due to a combination of what we are calling here cause 1 and cause 4 effects, would bring sensitivity to the expansion rate during recombination. Going a step further, Martins et al.~\cite{Martins:2010gu} studied correlated scaling of both $G$ and the fine-structure constant $\alpha$, allowing for a mechanism to leave the ratio of the photon scattering rate to Hubble rate unchanged. This allowed them to study new physics constraints in a model space where cause 1 had been approximately eliminated, allowing them to focus on what we identify here as cause 4. While all the observable consequences of cause 4 are ultimately due to its impact on the photon scattering rate, via small changes to the free-electron fraction $x_{\rm e}(z)$, we choose here to separate these recombination-related effects from cause 1. We have found it useful to distinguish this impact on the photon scattering rate relative to the Hubble rate from others, namely because even if we artificially fix recombination, changing the Hubble rate changes this rate ratio. 

By understanding better the constraints on light relics from CMB observations we have been able to construct light relics models that largely evade these constraints. With such models it is important to consider other sources of constraints. We thus include a brief discussion of constraints from ages of stars and from inferences of primordial light element abundances. 

This paper is organized as follows: In Sec.~\ref{sec:exact_symmetry} we present the scaling transformation from CGK and the origin of its associated symmetry in the Einstein-Boltzmann equations and initial conditions. In Sec.~\ref{sec:anal_models} we present power spectra that result from a series of approximations to the CGK scaling transformation. These scaling transformations are the ones available in particular model spaces, and hence our results here shed light on parameter constraints in these different model spaces. In Sec.~\ref{sec:constraints} we present and discuss parameter constraints in various light relics model spaces. 
We summarize and conclude in Sec.~\ref{sec:conclusions}.

\section{Symmetry of cosmological observations under scaling transformations} \label{sec:exact_symmetry}
In this section we show that the CMB anisotropy spectra are nearly invariant under the scaling transformation

\begin{align}\label{eq:scaling}
\sqrt{G\rho_i}  &\rightarrow \lambda  \sqrt{G\rho_i} , \nonumber \\
\sigma_{\rm T} n_{\rm e}  &\rightarrow  \lambda \sigma_{\rm T} n_{\rm e}, \\
A_{\rm s}  &\rightarrow  A_{\rm s}/\lambda^{(n_{\rm s}-1)},\nonumber
\end{align}
where $i$ enumerates all the components with densities $\rho_i$, and $A_{\rm s}$ and $n_{\rm s}$ are the amplitude and spectral index of the primordial power spectrum. The observables are exactly invariant if we (artificially) fix the ionization history $x_{\rm e}(z)$.  

To show the existence of this symmetry, we start in Sec.~IIA by computing how the solution to the Boltzmann equations describing cold dark matter (CDM), baryons, photons, and neutrinos transform under this scaling. We then discuss the behavior of the gravitational potentials under this transformation. In Sec.~IIB we show how the CMB power spectra can be left exactly invariant under this scaling once the primordial spectrum of fluctuations is properly adjusted. Our discussion follows that presented in Ref.~\cite{Zahn_2003}, but is extended here beyond the tight-coupling approximation. In Sec.~IIC we present a means of mimicking the gravitational free-fall rate scaling in a physical model. In Sec.~IID we discuss our implementation of photon scattering rate scaling and the dependence of the photon scattering rate on recombination rates relative to the Hubble rate. 

\subsection{Symmetry of the equations of motion}

As a starting point, let us first examine the  Boltzmann equations governing the evolution of photons and baryons fluctuations. Using the scale factor $a$ as our time variable, these take the form \citep{Ma:1995ey}
\begin{align}\label{eq:phot_bar}
   \frac{\pa F_{\gamma0}}{\pa a} & = - \frac{k }{ a^2 H}F_{\gamma 1} + 4\frac{\pa \phi}{\pa a} ,\\ 
    a^2 H\frac{\pa F_{\gamma1}}{\pa a} & = \frac{k}{3}(F_{\gamma0} - 2F_{\gamma2}) + \frac{4k}{3} \psi + \dot{\kappa}( \frac{4}{3}v_{\rm b}-F_{\gamma1}), \en
   a^2H\frac{\pa F_{\gamma2}}{\pa a} & = \frac{k}{5}(2F_{\gamma1} -3F_{\gamma3}) -\frac{9}{10}\dot{\kappa}F_{\gamma 2},\en
   a^2 H \frac{\pa F_{\gamma l}}{\pa a} & = \frac{k}{2l+1}\left[l F_{\gamma(l-1)} - (l+1)F_{\gamma(l+1)}\right] - \dot{\kappa}F_{\gamma l},\en
   \frac{\pa \delta_{\rm b}}{\pa a} & = - \frac{k}{a^2 H} v_{\rm b} + 3 \frac{\pa \phi}{\pa a}, \en
   a^2 H \frac{\pa v_{\rm b}}{\pa a} & = - a H v_{\rm b} + c_{\rm s}^2 k \delta_{\rm b} +  k\psi + \frac{\bar{\rho}_\gamma }{\bar{\rho}_{\rm b}} \dot{\kappa}(F_{\gamma1} -  \frac{4}{3}v_{\rm b}),\nonumber
\end{align}
where $F_{\gamma l}$ are the multipole moments of the photon temperature perturbation, $k$ is the Fourier wave number, $\dot{\kappa} = a\sigma_{\rm T} n_{\rm e} $ is the Thomson opacity, $\delta_{\rm b}$ is the baryon density perturbation, $v_{\rm b}$ is the baryonic bulk velocity, $c_{\rm s}$ is the baryonic sound speed, and $\phi$ and $\psi$ are the two gravitational potentials in the conformal Newtonian gauge. Note that we have used the relationship 
\begin{equation}
    \frac{d}{d\eta} = a^2 H \frac{d}{da}
\end{equation}
to convert between conformal time ($\eta$) derivatives and scale-factor derivatives. It is straightforward to see that these equations are invariant under the transformation
\begin{equation}\label{eq:key_transf}
    H\to \lambda H, \, k\to \lambda k,\, \sigma_{\rm T}n_{\rm e}\to \lambda \sigma_{\rm T}n_{\rm e}.
\end{equation}
These transformations correspond to equally rescaling \emph{all} length scales appearing in the Boltzmann equations: the Hubble horizon, the wavelength of fluctuations, and the photon mean free path.
 To close this system of equations, we need the perturbed Einstein equations for the $\phi$ and $\psi$ potentials. We use here the Poisson and shear equations
\begin{align}\label{eq:einstein}
    k^2\phi + 3 a H \left(a^2H\frac{d\phi}{da} + a H \psi\right) &= -4\pi G a^2 \sum_i \rho_i \delta_i, \\
    k^2(\phi - \psi) &= 12\pi G a^2 \sum_i (\rho_i + P_i)\sigma_i,\nonumber
\end{align}
where $\delta_i$, $\sigma_i$ and $P_i$ are the energy density perturbation, anisotropic stress and pressure of species $i$, respectively. These equations are invariant under the transformation given in Eq.~\eqref{eq:key_transf}, provided that the energy density of each component is individually rescaled, that is, $\sqrt{G\rho_i} \to \lambda \sqrt{G\rho_i}$. Massless neutrinos and dark matter follow collisionless versions of the equation given in Eqs.~\eqref{eq:phot_bar}, implying that they too are invariant under the transformation $H\to \lambda H$ and $k\to \lambda k$. We note that the evolution of massive neutrinos perturbations are also invariant under this transformation, once their masses are also properly rescaled.

We thus see that the linear evolution equations of \emph{all} components present in the Universe are invariant under the transformation
\begin{equation}\label{eq:key_transf2}
    \sqrt{G\rho_i} \to \lambda \sqrt{G\rho_i}, \, k\to \lambda k,\, \sigma_{\rm T}n_{\rm e}\to \lambda \sigma_{\rm T}n_{\rm e}.
\end{equation}
As we shall see in the next subsection, the rescaling of the wave number $k$ is actually unnecessary since one can express the solution $\tilde{\Phi}(k,a,\lambda)$ to the perturbation equations for a given wave number $k$ in the presence of the scaling in terms of the original solution in the absence of scaling but for a different wave number $k'=k/\lambda$
\begin{equation}\label{eq:scale_inv}
    \tilde{\Phi}(k,a,\lambda) = \Phi(k/\lambda,a,\lambda=1),
\end{equation}
where $\Phi,\tilde{\Phi}$ here stand for any of the perturbation variables (e.g.~$\delta,v, F_{\gamma l}$, etc.), and where it is now understood that both $\sqrt{G\rho_i} $ and $\sigma_{\rm T}n_{\rm e}$ are rescaled in the transformation. Such a relation was first presented in Ref.~\cite{Zahn_2003} in the context of the tight-coupling approximation ($\dot{\kappa} \gg H$), but we see here that it applies in a broader context once the Thomson opacity is also rescaled.

\subsection{Symmetry of the power spectra}
Having established the symmetry structure of the perturbed Boltzmann and Einstein equations, we now turn our attention to the symmetry of the actual observables: the power spectra. Our goal is to understand how the different power spectra transform (if at all) under the scaling transformation given in Eq.~\eqref{eq:key_transf2}. Here, we will focus on the CMB temperature power spectrum for simplicity, but note that the polarization and cross temperature-polarization spectra behave the exact same way under the scaling symmetry.  Under the scaling transformation, the CMB temperature power spectra can be written as
\be
C^{ TT}_\ell(\lambda) = \int \frac{dk}{k}P(k) |\tilde{\Delta}_{T\ell}(k,\lambda)|^2,
\ee
where $P(k)$ is the primordial spectrum of fluctuations, and $\Delta_{T\ell}(k,\lambda)$ is the photon transfer function under the transformation given in Eq.~\eqref{eq:key_transf2}. The latter can be written as 
\be
\tilde{\Delta}_{T\ell}(k,\lambda) = \int_0^1 da\, \tilde{S}_T(k,a,\lambda)j_\ell(k \tilde{\chi}(a,\lambda)), 
\ee
where $\tilde{S}_T(k,a,\lambda)$ is the photon source term, and $\tilde{\chi}(a,\lambda)$ is the comoving distance to \textcolor{red}{the} scale factor $a$ in the presence of the rescaled Hubble rate. The source term $\tilde{S}_T$ depends on cosmological perturbations obeying Eq.~\eqref{eq:scale_inv} and on the photon visibility function $\tilde{g}(a,\lambda) = -d/da(e^{-\tilde{\kappa}(a,\lambda)})$, where
\be
\tilde{\kappa}(a,\lambda) = \int_a^1 da' \frac{  \lambda \sigma_{\rm T} n_{\rm e}  }{  a' \lambda H} = \kappa(a,\lambda=1),
\ee
hence showing that the visibility function, once expressed in terms of the scale factor $a$, is invariant under the scaling transformation. This then ensures that 
\be
\tilde{S}_T(k,a,\lambda) = S_T(k/\lambda,a,\lambda=1).
\ee
Similarly, we have $\tilde{\chi}(a,\lambda) = \chi(a,\lambda=1)/\lambda$. We thus get
\begin{align}
\tilde{\Delta}_{T\ell}(k,\lambda) &= \int_0^1 da\, S_T(k/\lambda,a,\lambda=1)\en
&\qquad\qquad\qquad \times j_\ell((k/\lambda) \chi(a,\lambda=1))\en
& = \Delta_{T\ell}(k/\lambda,\lambda=1),
\end{align}
and the CMB temperature spectrum takes the form
\begin{align}\label{eq:cltt}
C^{ TT}_\ell(\lambda) &= \int \frac{dk}{k}P(k) |\Delta_{T\ell}(k/\lambda,\lambda=1)|^2\en
&=\int \frac{dk'}{k'} P(\lambda k') |\Delta_{T\ell}(k',\lambda=1)|^2.
\end{align}
Adopting the standard power-law primordial power spectrum $P(k) = A_{\rm s} (k/k_{\rm p})^{n_{\rm s}-1}$, where $k_{\rm p}$ is the pivot scale, and rescaling the scalar amplitude $A_{\rm s} \to A_{\rm s}/\lambda^{n_{\rm s}-1}$ we can write this as 
\begin{align}
C^{ TT}_\ell(\lambda) &= \int \frac{dk'}{k'} \frac{A_{\rm s}}{\lambda^{n_{\rm s}-1}} \left(\frac{\lambda k'}{k_{\rm p}}\right)^{n_{\rm s} -1} |\Delta_{T\ell}(k',\lambda=1)|^2\en
& = \int \frac{dk'}{k'} A_{\rm s} \left(\frac{ k'}{k_{\rm p}}\right)^{n_{\rm s} -1} |\Delta_{T\ell}(k',\lambda=1)|^2\en
& = C_\ell^{TT}(\lambda=1),
\end{align}
hence showing that the CMB temperature spectrum is indeed \emph{exactly} invariant under the transformation
\be
\left\{ \sqrt{G\rho_i}\to \lambda \sqrt{G\rho_i},\, \sigma_{\rm T}n_{\rm e} \to \lambda \sigma_{\rm T}n_{\rm e}, A_{\rm s} \to A_{\rm s}/\lambda^{n_{\rm s}-1}\right\}.
\ee
An entirely similar argument applies to the polarization and cross spectra, implying that the primary CMB is entirely unchanged under this transformation. Furthermore, matter clustering observables (e.g.~$\sigma_8$) are also invariant under this transformation (following a very similar argument to that presented above), which together with the invariance of distance ratios also leaves the lensing of the CMB unchanged. We thus conclude that the observable CMB is completely invariant under the above transformation. In what follows, we shall refer to this scaling as the Free-fall, Amplitude, and Thomson (FFAT) scaling.

Although our focus in this paper is on CMB power spectra, the invariance of BAO observables, cosmic shear power spectra, and matter power spectra under the FFAT transformation can be easily demonstrated with the arguments we have presented in this and the previous subsection. 

\subsection{Mirror world mimic of gravitational rate scaling}
\label{sec:MirrorWorld1}

In our Universe, the invariance under the FFAT scaling transformation is broken by direct measurements of the energy density of certain cosmological components. In particular, the FIRAS \cite{Fixsen1996,fixsen09} measurements of the mean temperature of the CMB photons today tightly constrain the photon energy density, $\rho_{\gamma,0}$, hence preventing the scaling transformation $\sqrt{G\rho_i}\rightarrow \lambda \sqrt{G\rho_i}$ from being implemented. To nonetheless exploit the scaling symmetry while maintaining agreement with our knowledge of the CMB thermal history, it is useful to consider a practical cosmological model containing a mirror world dark sector (MWDS) (see e.g.~Refs.~\cite{Chacko:2005pe,Chacko:2005vw,Chacko:2005un,Barbieri:2005ri,Craig:2013fga,Craig:2015pha,GarciaGarcia:2015fol,Craig:2015xla,Farina:2015uea,Farina:2016ndq,Prilepina:2016rlq,Barbieri:2016zxn,Craig:2016lyx,Berger:2016vxi,Chacko:2016hvu,Csaki:2017spo,Chacko:2018vss,Elor:2018xku,Hochberg:2018vdo,Francis:2018xjd,Harigaya:2019shz,Ibe:2019ena,Dunsky:2019upk,Csaki:2019qgb,Koren:2019iuv,Terning:2019hgj,Johns:2020rtp,Roux:2020wkp,Ritter:2021hgu,Curtin:2021alk,Curtin:2021spx}). As discussed in CGK, such a model can very closely mimic the $\sqrt{G\rho_i}\rightarrow \lambda \sqrt{G\rho_i}$ transformation while leaving the tightly constrained visible radiation and matter budget of the Universe unchanged. In this work, our main interest is not in the MWDS as a plausible cosmological scenario (see however Ref.~\cite{Blinov:2021mdk}), but rather as a useful reference model that can help us understand the origins of constraints on light relics. With this in mind, we briefly summarize the properties of the mirror sector that are most relevant to our work below. 

The MWDS contains a copy of baryons, photons and neutrinos in the dark sector. In this model, the relative energy density ratios of the dark baryon, dark photons and dark neutrinos are exactly the same as those in the visible sector. The presence of the dark photons effectively scales the photon gravitational free-fall rate while respecting the FIRAS constraint on the mean temperature of CMB photons. The addition of dark baryons acts to increase the energy density in baryonlike matter (i.e.~matter that cannot cluster prior to the epoch of recombination) while keeping the photon-to-baryon ratio unchanged in the visible sector. The addition of dark neutrinos, or any other free-streaming particles, are used to scale up the energy density of free-streaming species while preserving the free-streaming fraction of relativistic particles. That is, instead of scaling every component, we alter the scaling transformation so as to fix $\rho_{\rm b}$, $\rho_\gamma$, and $\rho_\nu$ while $\rho_{x}+ \rho_{x}^{\rm D} \rightarrow \lambda^2 \left( \rho_{x}+ \rho_{x}^{\rm D}\right)$ where $x$ indicates baryons, photons, or neutrinos and the superscript D denotes their dark sector versions.  

Within the dark sector, dark baryons (including dark electrons) and dark photons can interact via a Thomson-scattering like interaction. This interaction ensures that the fraction of the total matter density that can cluster is kept fixed at all times, as well as maintaining a similar pressure strength in the photon-baryon fluid in both visible and dark sectors. To ensure the photon perturbations of both sectors are in phase and that photon-baryon decoupling occurs at the same time in both sectors, we fix the ratio of hydrogen binding energy to photon temperature and keep the fine structure constant and proton mass the same in both sectors. We neglect any mixing between the dark and visible sectors, and assume that two sectors only interact through gravity. After dark recombination, the MWDS forms atomic dark matter  
\cite{Goldberg:1986nk,Fargion:2005ep,Khlopov:2005ew,Khlopov:2008ty,Kaplan:2009de,Khlopov:2010pq,Kaplan:2011yj,Khlopov:2011tn,Behbahani:2010xa,Cline:2012is,Cyr-Racine:2013ab,Cline:2013pca,Cyr-Racine:2013fsa,Fan:2013tia,Fan:2013yva,McCullough:2013jma,Randall:2014kta,Khlopov:2014bia,Pearce:2015zca,Choquette:2015mca,Petraki:2014uza,Cirelli:2016rnw,Petraki:2016cnz,Curtin:2020tkm}, which contributes a small fraction of the overall dark matter budget with the rest being made of standard CDM.

\subsection{Photon scattering and recombination rates scaling}

The second crucial transformation of the FFAT scaling [Eq.~\eqref{eq:scaling}] is the photon scattering rate $\sigma_{\rm T} n_{\rm e}$.
In this paper, this scaling is implemented by altering the primordial helium abundance, $Y_{\rm P}$. Since $n_{\rm e} \propto x_{\rm e}(1-Y_{\rm P})\rho_{\rm b}$, it is possible to change $Y_{\rm P}$ to enforce the photon scattering rate scaling [at fixed $x_{\rm e}(z)$] by performing the transformation
\begin{equation}\label{eq:yp_scaling}
    1-Y_{\rm P} \rightarrow \lambda (1-Y_{\rm P}),
\end{equation}
with $\rho_{\rm b}$ held fixed to its base \lcdm\ value due to the sensitivity of the CMB power spectra to the baryon-to-photon ratio. Much like the MWDS, we do not consider this scaling of the helium abundance away from its Big Bang Nucleosynthesis (BBN) prediction as part of a plausible cosmological scenario. Instead, we use this scaling of $Y_{\rm P}$ as a tool to understand light relics constraints from the CMB. See Sec.~\ref{sec:cosntraint_agebbn} below for a discussion of light-element abundances in the presence of the FFAT scaling. 

This scaling could also be implemented by rescaling the Thomson cross section $\sigma_{\rm T}\propto \alpha^2/m_{\rm e}^2$, where $\alpha$ is the fine-structure constant and $m_{\rm e}$ is the electron mass. This would require varying these fundamental constants (see e.g.~Refs.~\cite{Martins:2010gu,Sekiguichi:2021,Hart:2020,Hart:2021kad,Hart:2022agu,Burgess:2021qti,Burgess:2021obw}) away from their Standard Model values. Besides the obvious model-building challenge that this brings, care must also be taken if this route is chosen as other combinations of $\alpha$ and $m_{\rm e}$ enter the problem, most notably the Rydberg constant $\epsilon_0 \propto \alpha^2 m_{\rm e}$. While we do not follow this path here, we note that the scaling $\alpha \to \lambda^{1/6} \alpha$ and $m_{\rm e} \to \lambda^{-1/3} m_{\rm e}$ would achieve the right scaling of the Thomson cross section while leaving the Rydberg constant invariant. 

Since they are similar electromagnetic processes, the leading-order hydrogen recombination rate \cite{spitzer_book,Seager:1999km} and the photon scattering rate have the same parametric dependence on $\alpha$, $m_{\rm e}$, and $n_{\rm e}$, with both rates $\propto \sigma_{\rm T} n_{\rm e}$ (assuming the Rydberg constant is left unchanged). This means that, at leading order, a rescaling of the photon scattering rate automatically results in the same scaling of the hydrogen recombination rate. Thus, the FFAT scaling leaves the ratio of the leading-order hydrogen recombination rate to the Hubble rate unchanged, hence leaving the ionization history of the Universe nearly invariant. 

The ionization history is however not \emph{exactly} invariant under FFAT scaling due to the relative importance of other atomic rates to the overall hydrogen recombination process, most notably the two-photon $2s-1s$ transition rate, and the Lyman-$\alpha$ photon escape rate \cite{Peebles:1968ja}. The ratios of these rates to the Hubble expansion rate are not invariant under the FFAT transformation, resulting in small changes to the ionization history. This in turns leads to small modifications to the Thomson scattering rate and therefore to CMB spectra.

We are now ready to fully motivate our distinction between causes 1 and 4. We could certainly choose to organize our enumeration of causes so that they are both counted as cause 1. For both of them, observational consequences arise entirely from the changes to the ratio of the photon scattering rate to the Hubble rate. We have found it useful though to make the distinction for two reasons. First, 
even at fixed $x_{\rm e}(z)$ changing the Hubble rate generally leads to changes to $\sigma_{\rm T}n_{\rm e}(z)/H(z)$, and this change is generally larger than the change just due to changes to $x_{\rm e}(z)$ alone. Note that our focus on the ratio of rates, rather than $\sigma_{\rm T} n_e(z)$ alone, is appropriate because it is the ratio that leads to observational consequences; a changing ratio is a departure from uniform scaling. Second, the helium scaling provides us with a means of eliminating what we call cause 1, without eliminating what we call cause 4. In the next section we will see the impact of cause 4 in models that differ in ways that eliminate the other three causes. 

We close this section with a caveat about our ordering of the causes according to the magnitude of their impact.
Rescaling the gravitational free-fall rates $\sqrt{G\rho_i}\to \lambda \sqrt{G\rho_i}$ without a corresponding scaling of the photon scattering rate (and thus of the leading-order recombination rate) would result in much larger changes to the ionization history $x_{\rm e}(z)$. In this case, cause 4, although still subdominant to cause 1, would no longer be subdominant as compared to causes 2 and 3.

\section{Analysis of Models}\label{sec:anal_models}

Our approach to understanding constraints on light relics models starts with understanding power spectra differences between a best-fit \lcdm\ model and a point in the light relics model space with increased light relic abundance. Power spectra differences tend to be smaller if the chosen point in the light relics model space is along the trajectory of a scaling transformation away from the best-fit \lcdm\ model, and thus we choose such a parameter-space location for our comparisons. 

\begin{figure}[tbh]
\centering
\includegraphics[trim={1cm 2.5cm 2cm 1cm},clip,width=0.99\columnwidth]{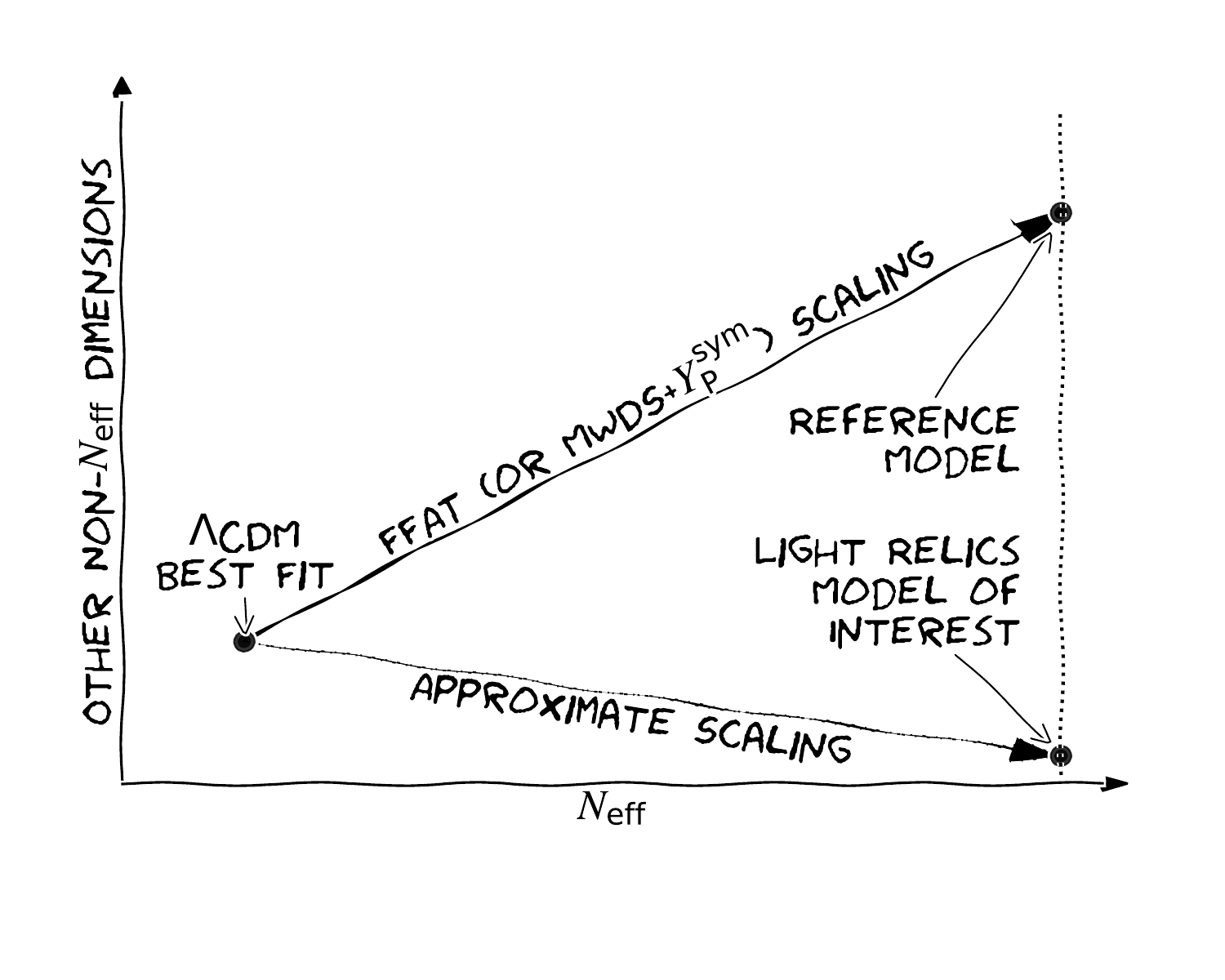}
\caption{A schematic visualization of the relationship in model space between a best-fit \lcdm\ model, the light relics (LR) model of interest, and a reference model used to understand the origin of the spectral differences between \lcdm\ and the LR model of interest. The reference model and the LR model have the same value of $N_{\rm eff}$ while the \lcdm\ model and the reference model have the same power spectra. Spectral differences between the LR model and the \lcdm\ model are thus the same as those between the LR model and the reference model. Model differences between the LR model and the reference model are fewer in number than are the model differences between the LR model and \lcdm\ and thus understanding the impact of these differences on spectra is easier. Spectral differences can be understood as arising from the constraints on the model space of interest that prevent it from following a FFAT scaling trajectory, and the impact of the resulting differences from FFAT scaling on the spectra.}
\label{fig:use_of_reference_model}
\end{figure}

To understand the origin of these power spectra differences, we replace the \lcdm\ model with a reference model that has the same (or very nearly the same) spectra as the best-fit \lcdm\ model and that has the same light relics abundance as the light relics model of interest. For a graphical summary of the relationship between these three models, see Fig.~\ref{fig:use_of_reference_model}. We use the MWDS plus free helium model space to define such a reference model. The advantage of this replacement is that we are now comparing two models that are much more similar; the differences between them are fewer in number and thus the origin of the spectral differences becomes clearer. At the same time, the replacement has not changed the power spectra differences, so we are still developing an understanding of the same differences. 

Thus motivated, in this section we examine the performance of various approximations to the FFAT scaling; i.e., we evaluate the changes that occur to CMB power spectra under these transformations. We begin with the most minor departures from the FFAT scaling, and proceed toward the more major departures. We thus start in Sec.~\ref{sec:MirrorWorld} with the MWDS, first with helium scaling and then without it. In Sec.~\ref{sec:MixAndADM} we include the helium scaling but drop the MWDS in favor of more vanilla additional light relics. The weakest departure from full scaling we study in this subsection preserves the free-streaming fraction of radiation (prior to hydrogen recombination) by using the appropriate mix of free-streaming and fluid additional light relics. We also study in Sec.~\ref{sec:RelicsAndYp} the two extremes of additional light relics that are purely free streaming or purely fluidlike. In Sec.~\ref{sec:vanilla} we look at the same model spaces, except with BBN-consistent helium instead. As we progress away from the FFAT scaling in this manner, new physical effects become important that drive differences with the \lcdm\ best-fit power spectra.  

All of the above cases are in model spaces with zero neutrino mass and the ionization history, $x_{\rm e}(z)$, fixed to a fiducial best-fit \lcdm\ value. In Sec.~\ref{sec:MnuAndPhysRec} we examine the quantitative symmetry-breaking impacts of the neutrino mass and out-of-equilibrium recombination. 

To fully define these scaling transformations we now point out where we are scaling from. We choose to scale from an \lcdm\ cosmology, which we refer to as the base $\lcdm$ model, with cosmological parameters\footnote{With the exception of $\Sigma m_\nu$ these are the parameter values for the best-fit Planck 2018 TTTEEE+LowE+lensing \lcdm\ cosmology \cite{Planck:2018vyg} with $\Sigma m_\nu$ fixed to 0.06 eV.} \footnote{We adopt $N_{\rm fs}=3.046$ from the Planck 2018 TTTEEE+LowE+lensing \lcdm\ cosmology in our analysis, but we note that recent calculations find  $N_{\rm eff} = 3.044$ as the expected value based on standard assumptions of the thermal history \cite{Akita:2020szl, Froustey:2020mcq, Bennett:2020zkv}. A change from 3.046 to 3.044 would have negligible impact on the our results.} $\{ H_0,\ \Omega_{\rm b} h^2,\ \Omega_ch^2,\ \tau,\ A_{\rm s},\  n_{\rm s},\ Y_{\rm P},\ N_{\rm fs},\ N_{\rm fld},\ \Sigma m_\nu\} = \{{67.36\ {\rm \left[ km/s/Mpc \right] }},\ 0.02237,\ 0.12,\ 0.0544,\ {2.1\times10^{-9}},\\ 0.9649,\ 0.2454,\ 3.046,\ 0,\ 0\}$.

The scaling transformations we examine are summarized in Table~\ref{tab:models}. They are arranged vertically so that as one moves down the page the degree of symmetry decreases, although there is some back and forth with respect to the photon scattering rate scaling. All of the transformations we consider have the $H(z) \rightarrow \lambda H(z)$ scaling, which in the model spaces we consider is guaranteed by $\rho_{\rm m}$, $\rho_{\rm rad}$, and $\rho_\Lambda$ all scaling up by $\lambda^2$. The case of FFAT scaling  
is the transformation given by Eq.~\eqref{eq:scaling}. Unless otherwise specified, the scalings we use in this section increase the Hubble parameter by 10\% (that is, $\lambda = 1.1$). For all scaling transformations listed in Table~\ref{tab:models}, the amplitude of the scalar fluctuations $A_{\rm s}$ is always adjusted according to $A_{\rm s}\to A_{\rm s}/\lambda^{n_{\rm s}-1}$.

Note that here we are strictly focused on scaling transformations and their impact on power spectra. In Sec.~\ref{sec:constraints} we will study the constraints on parameters in the model spaces that are related to these transformations. Power spectra changes, though useful for understanding parameter constraints, cannot be related directly to parameter constraints as degeneracies with other parameters can be important, and here we are not attempting any adjustments to maximize likelihoods.

\begin{table*}[htb]
 \centering
 \begin{tabular}{|c|c|c|c|c|c|}
 \hline
 {{{\multirow{2}{*}{\textbf{Scaling Transformation}}}}}&\multicolumn{5}{c|}{\textbf{Scaling Transformation Properties}}\\
 \cline{2-6}
 \multicolumn{1}{|c}{}&
 \multicolumn{1}{|c}{\begin{tabular}{@{}c@{}}$\rho_{\rm m} \propto \lambda^2$ \\ $\rho_{\rm rad} \propto \lambda^2$ \\ $\rho_\Lambda \propto \lambda^2$\end{tabular}}& 
 \multicolumn{1}{|c|}{$\sigma_T n_{\rm e} \propto \lambda$} &
 \multicolumn{1}{c|}{$R_{\rm fs}$ fixed} & 
 \multicolumn{1}{c|}{\begin{tabular}{@{}c@{}}$(\rho_\gamma + \rho_{\gamma}^D) \propto \lambda^2$ \\ $(\rho_{b}+\rho_{b}^D) \propto \lambda^2$ \end{tabular}}&
 \multicolumn{1}{|c|}{\begin{tabular}{@{}c@{}}$\rho_i \propto \lambda^2$ \\ $ \forall\ i$  \end{tabular}}
 \\
 \hline
 \multicolumn{1}{|c|}{Free-fall, Amplitude, and Thomson (FFAT) }&\multicolumn{1}{c|}{\cmark}&\multicolumn{1}{c|}{\cmark}&\multicolumn{1}{c|}{\cmark}&\multicolumn{1}{c|}{-}&\multicolumn{1}{c|}{\cmark}\\
 \multicolumn{1}{|c|}{Mirror World + $Y_{\rm P}^{\rm Sym}$(MWDS+ $Y_{\rm P}^{\rm Sym}$)}&\multicolumn{1}{c|}{\cmark}&\multicolumn{1}{c|}{\cmark}&\multicolumn{1}{c|}{\cmark}&\multicolumn{1}{c|}{\cmark}&\multicolumn{1}{c|}{\xmark}\\
 \multicolumn{1}{|c|}{Mirror World + $Y_{\rm P}^{\rm BBN}$(MWDS+ $Y_{\rm P}^{\rm BBN}$)}&\multicolumn{1}{c|}{\cmark}&\multicolumn{1}{c|}{\xmark}&\multicolumn{1}{c|}{\cmark}&\multicolumn{1}{c|}{\cmark}&\multicolumn{1}{c|}{\xmark}\\
 \multicolumn{1}{|c|}{$\Delta N_{\rm fs}+\Delta N_{\rm fld}+Y_{\rm P}^{\rm Sym}$ (Mix$+Y_{\rm P}^{\rm Sym}$)}&\multicolumn{1}{c|}{\cmark}&\multicolumn{1}{c|}{\cmark}&\multicolumn{1}{c|}{\cmark}&\multicolumn{1}{c|}{-}&\multicolumn{1}{c|}{\xmark}\\
  \multicolumn{1}{|c|}{$\Delta N_{\rm fld}+Y_{\rm P}^{\rm Sym}$}&\multicolumn{1}{c|}{\cmark}&\multicolumn{1}{c|}{\cmark}&\multicolumn{1}{c|}{\xmark}&\multicolumn{1}{c|}{-}&\multicolumn{1}{c|}{\xmark}\\
  \multicolumn{1}{|c|}{$\Delta N_{\rm fs}+Y_{\rm P}^{\rm Sym}$}&\multicolumn{1}{c|}{\cmark}&\multicolumn{1}{c|}{\cmark}&\multicolumn{1}{c|}{\xmark}&\multicolumn{1}{c|}{-}&\multicolumn{1}{c|}{\xmark}\\
  \multicolumn{1}{|c|}{$\Delta N_{\rm fs}+\Delta N_{\rm fld}+Y_{\rm P}^{\rm BBN}$ (Mix$+Y_{\rm P}^{\rm BBN}$)}&\multicolumn{1}{c|}{\cmark}&\multicolumn{1}{c|}{\xmark}&\multicolumn{1}{c|}{\cmark}&\multicolumn{1}{c|}{-}&\multicolumn{1}{c|}{\xmark}\\
 \multicolumn{1}{|c|}{$\Delta N_{\rm fld}+Y_{\rm P}^{\rm BBN}$}&\multicolumn{1}{c|}{\cmark}&\multicolumn{1}{c|}{\xmark}&\multicolumn{1}{c|}{\xmark}&\multicolumn{1}{c|}{-}&\multicolumn{1}{c|}{\xmark}\\
 \multicolumn{1}{|c|}{$\Delta N_{\rm fs}+Y_{\rm P}^{\rm BBN}$}&\multicolumn{1}{c|}{\cmark}&\multicolumn{1}{c|}{\xmark}&\multicolumn{1}{c|}{\xmark}&\multicolumn{1}{c|}{-}&\multicolumn{1}{c|}{\xmark}\\
 \hline
\end{tabular}
\caption{Properties of the scaling transformations applied in this section. A check mark (\cmark) means the corresponding scaling transformation property has been implemented, while a cross (\xmark) means it has not. The fraction of radiation energy density in free-streaming species is denoted as $R_{\rm fs}$. The superscript ``D" indicates a component in the MWDS discussed in Sec.~\ref{sec:MirrorWorld}. Transformations in model spaces without an MWDS get a `-' in the second-from-right column indicating that property is not applicable. FFAT scaling means scaling according to Eq.~\eqref{eq:scaling}. Note that all of these scaling transformations implicitly include $A_{\rm s} \rightarrow A_{\rm s}/\lambda^{(n_{\rm s}-1)}$ and, except when noted otherwise, an artificially fixed recombination history $x_{\rm e}(z)$.
}
\label{tab:models}
\end{table*}

\subsection{Mirror world dark sector}
\label{sec:MirrorWorld}

\begin{figure*}[tbh]
\centering
\includegraphics[trim={4cm 2.5cm 4cm 0cm},clip,width=\textwidth]{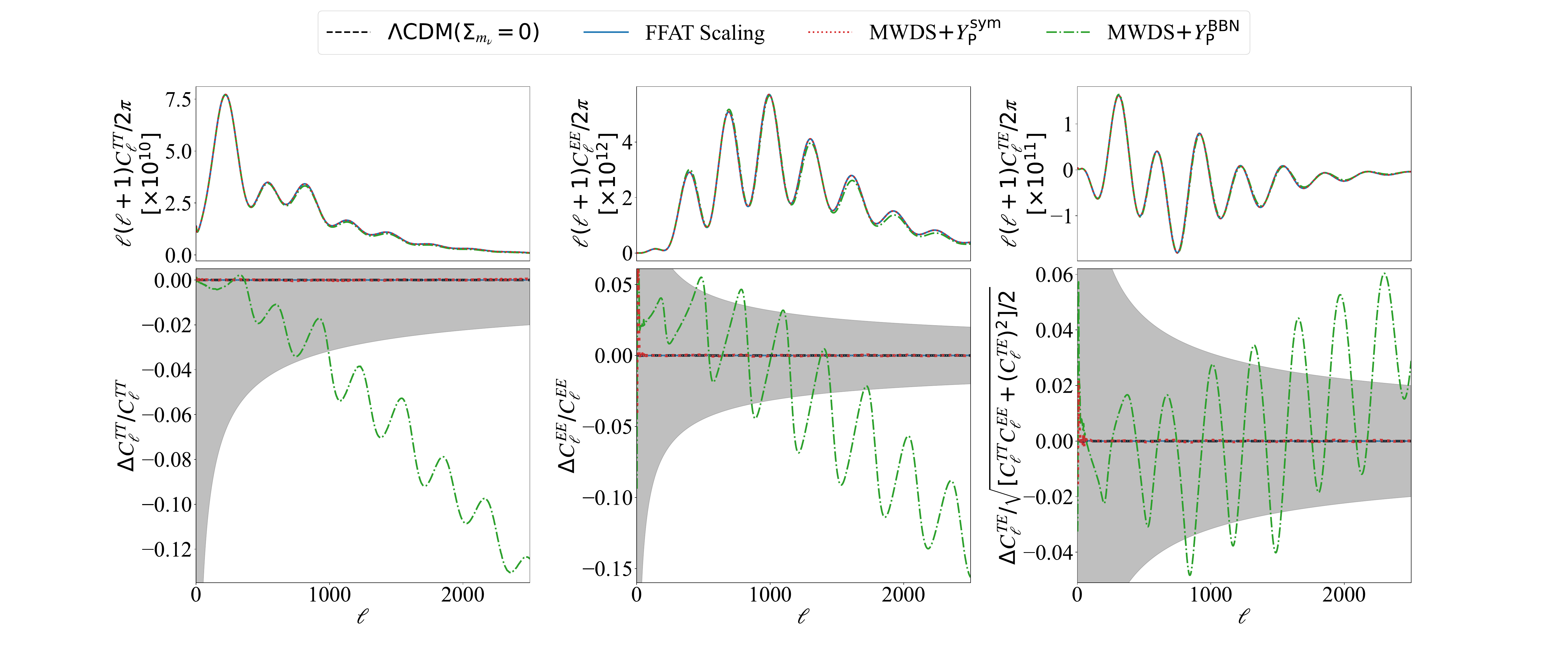}
\caption{CMB power spectra comparisons of the best-fit $\lcdm\ (\Sigma m_\nu=0)$ model to models scaled from there via FFAT scaling and MWDS scaling by $\lambda = 1.1$ to $H_0=74.1$ km/s/Mpc. The black dashed lines show the base $\lcdm\ (\Sigma m_\nu=0)$ model. The blue solid lines show the CMB power spectra after FFAT scaling. The symbol $Y_{\rm P}^{\rm BBN}$ means that the primordial helium fraction is derived from the BBN prediction, while $Y_{\rm P}^{\rm Sym}$ means the helium abundances is set by the scaling transformation of Eq.~\eqref{eq:yp_scaling}. All of the fractional differences in the bottom panels are compared to the $\lcdm\ (\Sigma m_\nu=0)$ case. Cosmic variance in individual multipoles is shown as a gray band.}
\label{fig:model_mirror_world}
\end{figure*}

Motivated by the scaling transformation in Sec.~\ref{sec:exact_symmetry}, we investigate the FFAT scaling model and its practical implementation using the MWDS. The CMB power spectra are shown in Fig.~\ref{fig:model_mirror_world}. 

~First, note that the CMB angular power spectra of the FFAT scaling model (blue lines) following Eq.~\eqref{eq:scaling} are exactly the same as those in the base \lcdm\ model (black dashed lines). The CMB power spectra are indeed invariant with massless neutrinos and fixed ionization history. This is exactly as expected from Sec.~\ref{sec:exact_symmetry}, because the power spectra are dimensionless.

As pointed out in the previous section it is not possible to implement the FFAT scaling transformation without breaking the observational constraint on the CMB mean temperature today, but we can use a MWDS to mimic it. In Fig.~\ref{fig:model_mirror_world}, we show the power spectra of the MWDS models maximally exploiting the FFAT scaling in red dotted lines (MWDS+$Y_{\rm P}^{\rm Sym}$ model).  We see that the MWDS scaling transformation together with the helium abundance reduction to boost the photon scattering rate leads to CMB power spectra that are very nearly identical to those of FFAT scaling, and therefore of \lcdm. 

In contrast, if one instead chooses to not scale the photon scattering rate, the power spectra change significantly. In Fig.~\ref{fig:model_mirror_world}, the power spectra of the MWDS+$Y_{\rm P}^{\rm BBN}$ model (green lines) are obviously different from the other spectra shown in the plot. In this model, the photon scattering rate is not scaled, but calculated following the BBN-predicted $Y_{\rm P}$. The TT spectrum gets smaller with $\ell$ getting larger compared to the other models in the figure. The EE spectrum is larger at low-$\ell$ and smaller in the high-$\ell$ tail. 

These changes to the spectra are all due to the photon-scattering rate being below its FFAT scaling value, a consequence of our use of a BBN-consistent $Y_{\rm P}$. In the language of our four causes, this is cause 1: a change to the photon scattering rate relative to the Hubble rate [at artificially fixed $x_{\rm e}(z)$]. Due to the increased expansion rate, there is an increased helium yield from BBN, due to both neutron-proton freeze-out at higher temperature and less time for neutron decay between then and the onset of helium production \citep{Kolb:1990vq}.  This is the opposite direction of change in helium yield that we would want to appropriately scale the photon scattering rate; at a given baryon density, the boosted helium leads to a reduction in free electrons density. One result is an increased photon mean free path during the recombination period and thus increased diffusion damping.  The last scattering period is also longer due to the larger photon mean free path, leaving a longer time for quadruple moments of the photon perturbations to grow \citep{Zahn_2003}. As the main source of polarization \citep{Zaldarriaga:1995gi}, the larger quadruple moments boost polarization amplitude. At small scales, the polarization amplitude is damped due to photon diffusion. Thus, the boosted amplitude is only seen at large scales.

The net result of this isolated cause 1 effect with $\lambda = 1.1$ (which brings us to $N_{\rm eff} = 4.61$) are spectral differences at the 10 to 15\% level. 

\subsection{Free streaming and fluid light relics with scattering rate scaling}
\label{sec:MixAndADM}

\begin{figure*}[tbh]
\centering
\includegraphics[trim={4cm 2.5cm 4cm 0cm},clip,width=\textwidth]{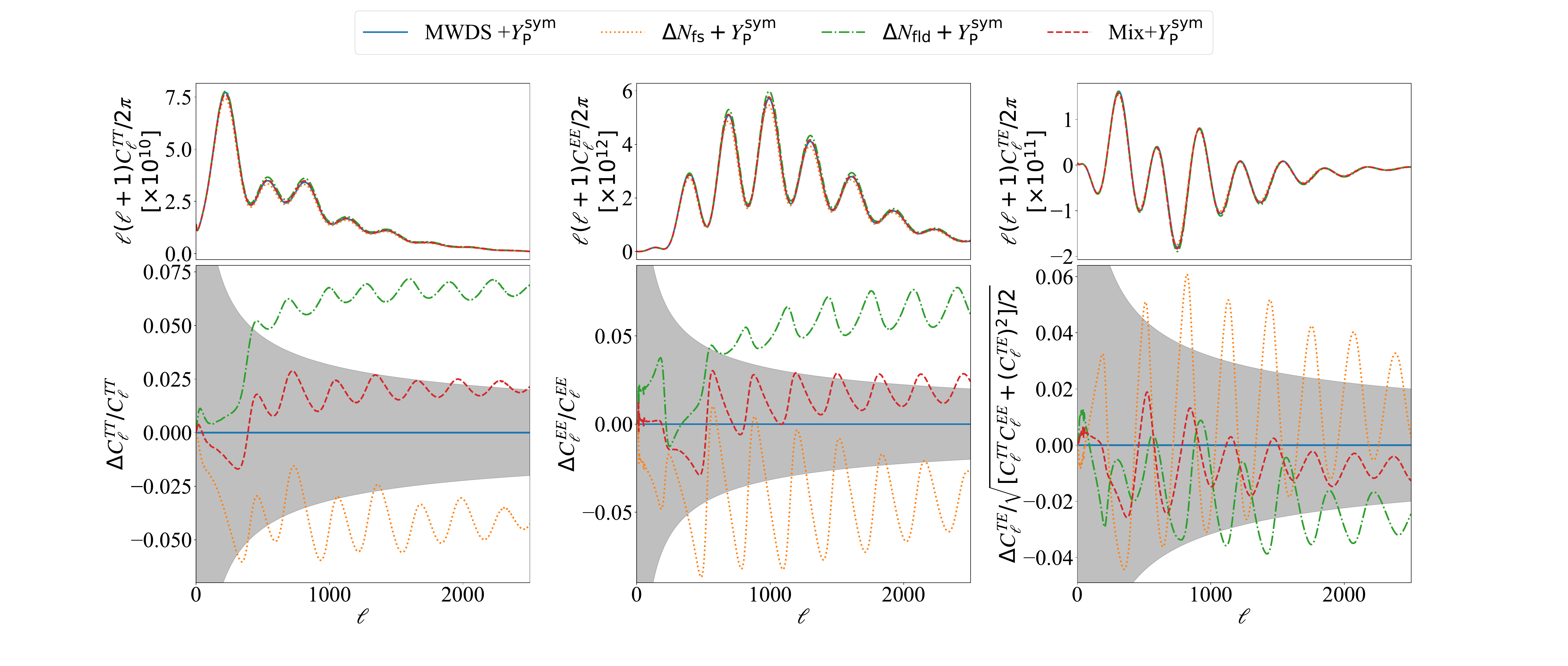}
\caption{CMB power spectra comparisons of the MWDS+$Y_{\rm P}^{\rm Sym}$ model to several other light relics models. In all models, $\lambda =1.1$. The increase in the effective number of neutrino species due to free-streaming (or fluidlike) species is $\Delta N_{\rm fs}$ ($\Delta N_{\rm fld}$). We use the MWDS+$Y_{\rm P}^{\rm Sym}$ model as the reference model, which is indistinguishable from the corresponding FFAT scaling model. The fractional differences in the bottom panels are relative to the reference model. Cosmic variance in individual multipoles is shown as gray bands.}
\label{fig:model_scale_yp}
\end{figure*}

Having established in the previous subsection that the MWDS can provide a highly effective mimic of the FFAT scaling transformation, we can now look at simpler models of light relics and understand the changes in power spectra as arising from the differences between these models and the MWDS model. Here, and in the next subsection, we keep the photon scattering rate scaling, and extend \lcdm\ with either free-streaming or fluid light relics or a mix of the two. These model spaces cannot mimic the scaling of all the gravitational free-fall rates $\sqrt{G\rho_i}$, but still allow for $H(z) \rightarrow \lambda H(z)$. 

It is in this subsection that we present our answer to the question that initially motivated the investigations that led to CGK and this paper. That question is, how do the CMB data lead to constraints on light relics models when one has addressed the damping scale and polarization generation problems by freeing up helium (eliminating cause 1), and eliminated impacts of a changing free-streaming fraction (cause 2)?

As is conventional we parametrize the energy density in free-streaming and fluidlike species in terms of their effective number of neutrino species, $N_{\rm fs}$ and  $N_{\rm fld}$ respectively so that the total radiation energy density is given by
\begin{equation}
    \rho_{\rm rad}(z) = \rho_{\gamma, 0}(1+z)^4\left[1+\left(N_{\rm fs}+N_{\rm fld}\right)\frac{7}{8}\left(\frac{4}{11}\right)^{4/3}\right],
\end{equation}
where $\rho_{\gamma,0}$ is the photon mean energy density today tightly constrained by FIRAS. In our base \lcdm\ model $(N_{\rm fs},N_{\rm fld}) = (3.046, 0)$.

To ensure $H(z) \rightarrow \lambda H(z)$ we have to send $\rho_\Lambda \rightarrow \lambda^2 \rho_\Lambda$, $\rho_{\rm m}(z) \rightarrow \lambda^2 \rho_{\rm m}(z)$, and $\rho_{\rm rad}(z) \rightarrow \lambda^2 \rho_{\rm rad}(z)$. For both matter and radiation there are choices to be made. For matter here we always choose to keep the baryon density fixed, in order to keep the baryon-to-photon ratio fixed. We increase $\rho_{\rm cdm}$ by a sufficient amount to ensure the desired $\rho_{\rm m}(z)$ scaling. For the radiation density, we make the minimum departure from MWDS scaling by having the appropriate mix of free-streaming and fluidlike additional relics to preserve the free-streaming fraction of radiation,
\begin{equation}
     R_{\rm fs} = \rho_{\rm fs}/\rho_{\rm rad},
\end{equation}
and refer to this model as $\Delta N_{\rm fs}+\Delta N_{\rm fld} + Y_{\rm P}^{\rm Sym}$ or Mix $+Y_{\rm P}^{\rm Sym}$. In Sec.~\ref{sec:RelicsAndYp}, we also consider the two extreme cases of increasing the radiation density with either free-streaming particles alone ($\Delta N_{\rm fs}+ Y_{\rm P}^{\rm Sym}$) or fluidlike additional relics alone ($\Delta N_{\rm fld} + Y_{\rm P}^{\rm Sym}$).

In Fig.~\ref{fig:model_scale_yp} we compare the power spectra of these three models to the MWDS+$Y_{\rm P}^{\rm Sym}$ spectra. Let us consider the free-streaming-ratio-preserving case (Mix+$Y_{\rm P}^{\rm Sym}$). We see for this model a slight decrease in power on large scales and a slight increase in power on smaller scales. We turn to the differences with the corresponding MWDS model for our explanation of these spectral differences. These models, prior to recombination, have the same free-streaming radiation densities, the same fluid radiation densities, and the same dark matter densities. They differ in two distinct ways: first, the atomic dark matter in the MWDS model is replaced with additional cold dark matter in the model of interest and second, the fluidlike dark radiation in the MWDS model transitions to free-streaming radiation at recombination (which, by design of our particular MWDS model, happens at the same time whether it is visible recombination or dark recombination).

{\em We find that the differences in the spectra are predominantly due to the lack of pressure support for the cold dark matter that would be dark baryons 
in the corresponding MWDS model, a lack of pressure support that alters gravitational potentials.} 

This finding, evidence for which we will present shortly, should not be surprising. The most dramatic difference between the two models is that what is atomic dark matter in one model, experiencing pressure support, is cold dark matter in the other model, experiencing none. The differences in the radiation content are much milder: although the dark photons effectively form a fluid with the dark baryons, with a somewhat reduced sound speed compared to the fluidlike light relics, the dark photons and fluidlike light relics both experience pressure support. After recombination the dark photons (by design) are free streaming, unlike the fluidlike light relics which remain fluidlike, but by this time the radiation is a small contributor to the total density. 

The cause of the differences between these spectra can also be expressed in a manner independent of the properties of the reference model: it is due to the change to the fraction, prior to recombination, of nonrelativistic matter that is pressure supported. This is our cause 3. In the Mix + $Y_{\rm P}$ model this fraction is simply $\rho_{\rm b}/\rho_{\rm m}$. In the MWDS + $Y_{\rm P}$ model it is $(\rho_{\rm b} + \rho_{\rm b}^{\rm D})/\rho_{\rm m}$. 

\begin{figure}[!]
\centering
\includegraphics[trim={0cm 0cm 0cm 0cm},clip,width=\columnwidth]{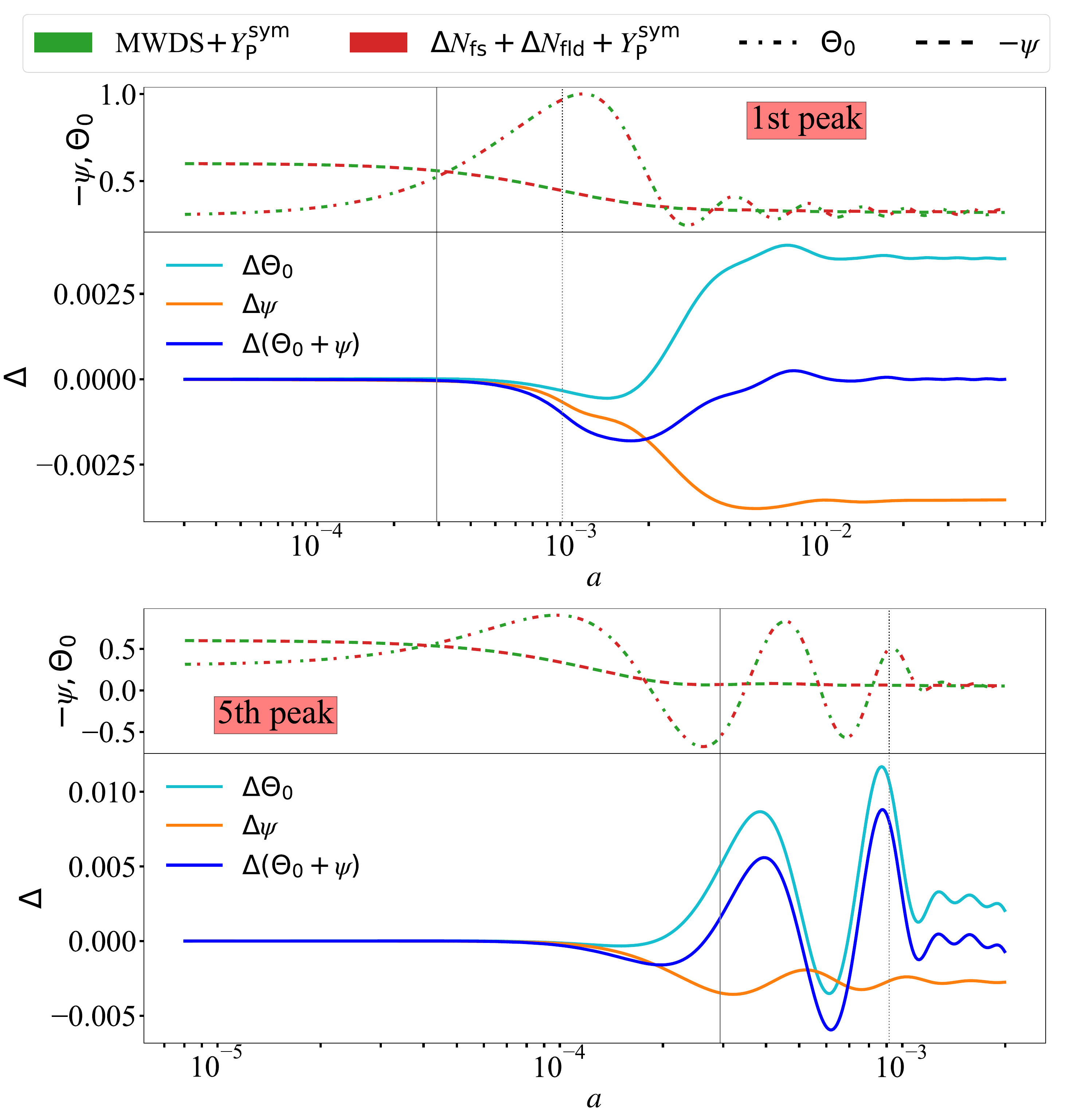}
\caption{Evolution of $\Theta_0$ and $\psi$ of the scales corresponding to the first and fifth peak in the temperature power spectrum of the Mix + $Y_{\rm P}^{\rm Sym}$ model (in green) and MWDS+$Y_{\rm P}^{\rm Sym}$ model (in red). The top parts of both panels show the amplitude of $\Theta_0$ (dash-dotted lines) and $-\psi$ (dashed lines). The bottom parts of both panels show the difference of $\Theta_0$ (light blue), $\psi$ (orange) and $\Theta_0+\psi$ (dark blue) of the Mix $+ Y_{\rm P}^{\rm Sym}$ model compared to the MWDS + $Y_{\rm P}^{\rm Sym}$ model; e.g., the light blue curve in the residual panel is $\Theta_0^{{\rm Mix} + Y_{\rm P}^{\rm Sym}} - \Theta_0^{{\rm MWDS} + Y_{\rm P}^{\rm Sym}}$. The solid and dotted vertical lines show the scale factors of matter-radiation equality and last scattering, respectively.}
\label{fig:theta_psi}
\end{figure}

Let us first look at what happens on larger scales by examining the evolution of the photon monopole perturbation moment, $\Theta_0$, and $\psi$ in our two models for a perturbation wavelength that contributes significantly to the first peak of the temperature power spectrum. We see in Fig.~\ref{fig:theta_psi} that in the Mix + $Y_{\rm P}^{\rm Sym}$ model that the gravitational potential decays less than in the MWDS model (since $\psi < 0$, the lesser amount of decay shows up in the residual plot as a negative $\Delta \psi$). This is expected since the pressure support in the MWDS model leads to greater potential decay. The smaller amount of decay reduces the resonant driving of the baryon-photon fluid oscillations and thereby reduces $\Theta_0$ at last scattering, at which time the oscillator of this mode reaches its first compression. The net result is an even larger decrease in the effective temperature $\Theta_0 + \psi$, which, since this is a positive quantity, is also a decrease in its amplitude. Thus we see a suppression in power near the first peak. 

On smaller scales things play out a bit differently as shown in the bottom panel of Fig.~\ref{fig:theta_psi}. There we show, as a representative case, the evolution of $\Theta_0$ and $\psi$ in our two models for a perturbation wavelength that contributes to the 5th peak of the temperature spectrum. With this smaller-wavelength mode, horizon crossing happens earlier when the radiation-to-matter ratio is higher. In both models, substantially more gravitational potential decay occurs and there is more scale-factor evolution between horizon crossing and last-scattering. Unlike with modes that contribute to the first peak, potential differences begin to emerge after the first compression, driving changes to the baryon-photon fluid that, by the time of last-scattering, boost the amplitude of both $\Theta_0$ and $\Theta_0 + \psi$.

We now identify the main factor leading to the gravitational potential differences. Comparing the two modes in Fig.~\ref{fig:theta_psi}, the residual changes are similar from the horizon entry until finishing the first maximum contraction. The difference starts to emerge during the sub-horizon evolution after the horizon entry until the last scattering. During this time, the growth of photon temperature monopole is halted due to the pressure within the photon-baryon fluid. The CDM perturbation is able to keep growing. As a result, the CDM perturbation will become the main source driving the potential growth. In turn, the photon perturbation multipoles are driven by the potential in Eq.~\eqref{eq:phot_bar}. To see this, we use the time-space component of the Einstein equation, 
\begin{equation*}
    k^2 \left( \dot{\phi}+\frac{\dot{a}}{a} \psi \right) = 4 \pi G a^2 \sum_i (\rho_i+P_i) \theta_i,
\end{equation*}
with Eq.~\eqref{eq:einstein} to get expressions for gravitational potential without time derivatives:
\begin{align}\label{eq:potential}
    k^2\phi = -4\pi G a^2 &\left[ \sum_i \rho_i \delta_i+\frac{3aH}{k^2}\sum_i (\rho_i + P_i) \theta_i\right], \\
    k^2 \psi = -4\pi G a^2 & \left[  \sum_i \rho_i \delta_i + \frac{3aH}{k^2}\sum_i (\rho_i + P_i) \theta_i \right.\nonumber\\ 
  &\left. \quad + 3 \sum_i (\rho_i + P_i)\sigma_i \right],\nonumber
\end{align}
where $\theta_i$ is the velocity divergence of specie $i$. The difference in the potential perturbation comes from the difference in the CDM density perturbation. Under the subhorizon limit,  $\frac{aH}{k}<<1$, the velocity term can be neglected compared to the other terms. Proportional to $R_{\rm fs}$ \citep{Bashinsky:2003tk, Baumann:2015rya}, the shear term has the same contribution between the Mix+$Y_{\rm P}^{\rm Sym}$ model and MWDS+$Y_{\rm P}^{\rm Sym}$ model, where the free-streaming fractions are the same. The density perturbation is dominated by the CDM, $\Sigma_i \rho_i \delta_i \approx \rho_c \delta_c$, at subhorizon scale.

Looking at the Boltzmann equations of Eq.~\eqref{eq:phot_bar}, the gravitational potential impacts the evolution of the photon perturbations. After horizon entry, the gravitational potential decays and oscillates. For the Mix+$Y_{\rm P}^{\rm Sym}$ model, it is the reduced gravitational potential decay post first compression that drives the photon perturbation to a larger amplitude. 
In the residual plot of the mode corresponding to the 5th peak of Fig.~\ref{fig:theta_psi}, we see the difference in $\Theta_0$ grows with time. At last scattering, $\Theta_0+\psi$ is at a larger amplitude in Mix+$Y_{\rm P}^{\rm Sym}$ model. As a result, the CMB power spectrum has larger amplitude at small scales, which we see as the excess power in the TT and EE residuals in Fig.~\ref{fig:model_scale_yp}.

A change in the gravitational potential amplitude also modifies the zero-point equilibrium position of the $\Theta_0+\psi$ oscillator. This results in alternating higher and lower peaks in the CMB temperature spectrum residual at small scales. While present, this pattern is partially obscured in the temperature residuals shown in Fig.~\ref{fig:model_scale_yp} by other out-of-phase source contributions (especially cross terms) to the total CMB spectrum. To better illustrate this point, we compare in Appendix~\ref{apd:cls} the different contributions to the spectral differences shown in Fig.~\ref{fig:model_scale_yp}, including the contributions from the monopole term $\Theta_0 + \psi$, the velocity of the plasma, the early Integrated Sachs-Wolfe Effect, and gravitational lensing. We find that the monopole term dominates over most of the higher $\ell$ range, with the boosted $\Theta_0$ and shift to the oscillator's zero-point clearly visible there. We also find that the eISW and the velocity term are both significant on larger scales. Lensing starts to make $\sim 10\% $-level changes to the fractional differences at $\ell > 2000$.

In this subsection, we have studied what we call the Mix+$Y_{\rm P}^{\rm Sym}$ model and addressed the question of why its CMB spectra differ from those of \lcdm\ even when we choose points in the respective model spaces that both have the same shape of $H(z)$, scattering rates, and free-streaming fractions (prior to recombination). Such a comparison has no cause 1 or cause 2 and because we also artificially fix $x_{\rm e}(z)$ there is also no cause 4. 

Our explanation relies on our use, as a reference model, of an MWDS + $Y_{\rm P}^{\rm Sym}$ model obtained by a scaling transformation away from the best-fit \lcdm\ model. We identified the important difference between the Mix + $Y_{\rm P}^{\rm Sym}$ model and the reference model as the fraction of CDM in the former model that is replaced with atomic dark matter in the latter model. This change is a change to the fraction of non-relativistic matter that is pressure supported; i.e., this is cause 3. The resulting difference in pressure support leads to differences in gravitational potentials, which in turn lead to differences in the CMB spectra. We see differences in the spectra, for $\lambda = 1.1$, are at the 2 to 3\% level. At least for the TT spectrum, for most values of $\ell$, these changes were driven mostly by changes to the monopole $\Theta_0 + \psi$ at recombination, rather than ISW effects, the Doppler term or gravitational lensing.

\subsection{Additional pure free-streaming or pure fluidlike light relics}
\label{sec:RelicsAndYp}

In the previous subsection we introduced three models and focused our attention on the one that preserves the free-streaming fraction, $R_{\rm fs}$. Now we turn to the other two models. The $H(z)\rightarrow \lambda H(z)$ transformation is still enforced in the models, but the additional light relics are free-streaming species only ($\Delta N_{\rm fs}+Y_{\rm P}^{\rm Sym}$) or fluidlike relics only ($\Delta N_{\rm fld}+Y_{\rm P}^{\rm Sym}$). The CMB power spectra are shown in Fig.~\ref{fig:model_scale_yp}. The main changes to the power spectra, compared to the Mix+$Y_{\rm P}^{\rm Sym}$ and MWDS+$Y_{\rm P}^{\rm Sym}$ models, are the overall amplitude difference and the temporal phase shift. These effects have been studied in Refs.~\citep{Bashinsky:2003tk, Baumann:2015rya}.

In the pure free-streaming model power is suppressed and in the pure fluid model power is enhanced. The free-streaming species reduce the super-horizon solution to the gravitational potential due to the shear induced by neutrinos, leading to a less ``radiation driving" boost to the photon perturbation after horizon entry. Besides, the neutrinos freeze out earlier than last scattering and start to free stream with a speed faster than the sound speed of the photon-baryon plasma. The neutrino perturbation and the photon perturbation of the same scale are out-of-resonance, reducing the photon perturbation amplitude. As a result, adding more free-streaming species will reduce the amplitude of the power. Also in these models there are temporal phase shifts to the acoustic oscillations that are proportional in amplitude to $R_{\rm fs}$, that lead to shifts in the peaks and therefore oscillatory features in the residual power spectra. These lead to especially pronounced oscillations in the residuals in the pure free-streaming case as their sign leads them to interfere constructively with the oscillations that are already there in the fixed $R_{\rm fs}$ case.
 
In the previous subsection we isolated effects from cause 3. In this one we see the combined effects of causes 2 and 3 for two different models. We can see from the residuals in Fig.~\ref{fig:model_scale_yp} that cause 2 and cause 3, for $\lambda = 1.1$, are leading to changes in spectra at the 5 to 6\% level. 
 
\subsection{Free streaming and fluid light relics without photon scattering rate scaling}
\label{sec:vanilla}

\begin{figure*}[!]
\centering
\includegraphics[trim={4cm 2.5cm 4cm 0cm},clip,width=\textwidth]{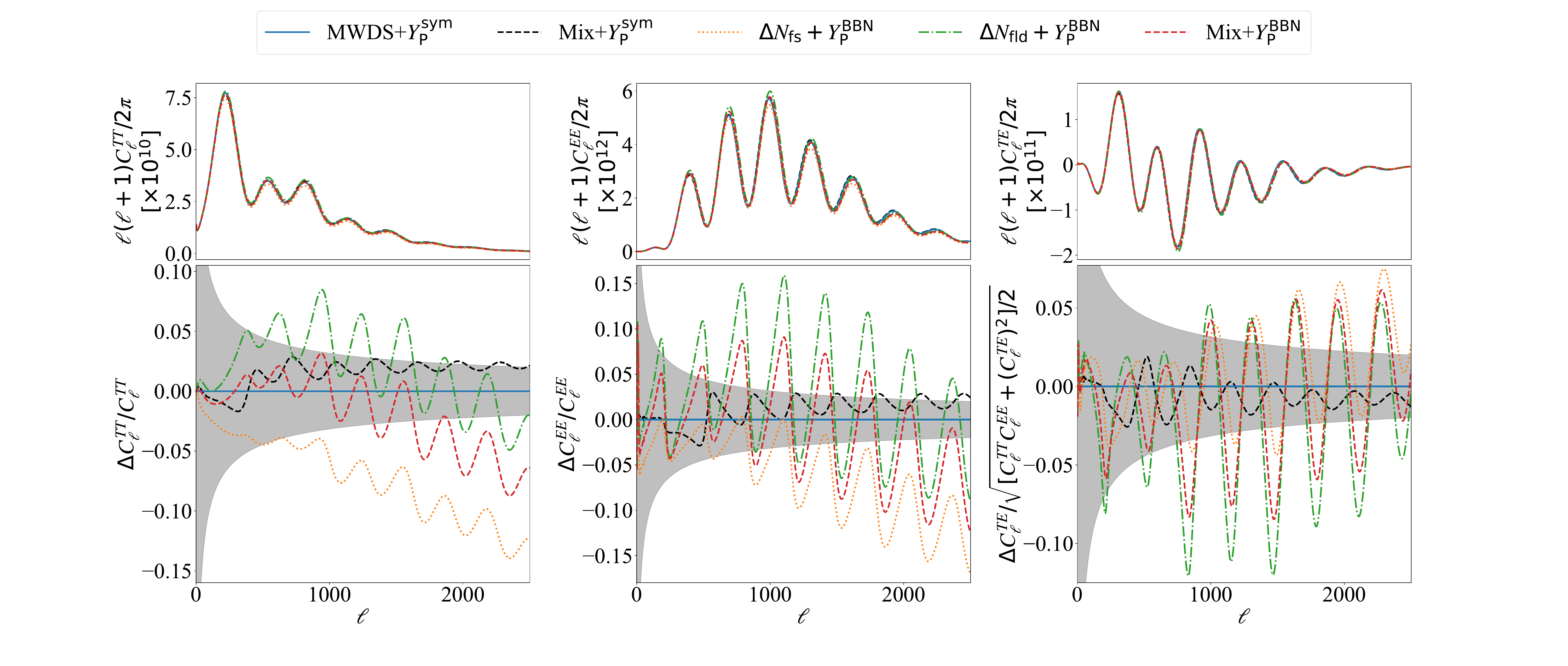}
\caption{CMB power spectra comparisons of free-streaming and fluid light relics models without scattering rate scaling to the Mix+$Y_{\rm P}^{\rm Sym}$ model and MWDS+$Y_{\rm P}^{\rm Sym}$ model. In all models, $\lambda=1.1$. The MWDS+$Y_{\rm P}^{\rm Sym}$ CMB power spectra are shown in solid blue lines.
The black dashed lines show the same spectra of the Mix+$Y_{\rm P}^{\rm Sym}$ model as in Fig.~\ref{fig:model_scale_yp}. The increase in the effective number of neutrino species due to free-streaming (or fluidlike) species is $\Delta N_{\rm fs}$ ($\Delta N_{\rm fld}$). By $Y_{\rm P}^{\rm BBN}$ we mean that the primordial helium fraction is derived from BBN prediction. All of the fractional differences in the bottom panels are relative to the reference model, MWDS+$Y_{\rm P}^{\rm Sym}$. Cosmic variance in individual multipoles is shown as gray bands.}
\label{fig:model_vanilla}
\end{figure*}

In this subsection, we continue our movement away from FFAT scaling by looking at the models of Sec.~\ref{sec:MixAndADM}\ and \ref{sec:RelicsAndYp}, but now with BBN-consistent $Y_{\rm P}$ instead of enforcing the primordial helium abundance as in Eq.~\eqref{eq:yp_scaling}. This effectively removes that scaling of the photon scattering rate. In Fig.~\ref{fig:model_vanilla}, we compare the power spectra resulting from these three different scaling transformation with the Mix+$Y_{\rm P}^{\rm Sym}$ model and with the reference model. We see in these residuals the combined effects of causes 1, 2, and 3. 
The impacts of cause 1 can be seen in the decreasing amplitude in power at high-$\ell$ tail and the significant changes in the EE spectra residuals. The overall power spectrum amplitude differences of the three models are induced by cause 2. The impact of cause 3 can only be seen in the region around the first two peaks between Mix+$Y_{\rm P}^{\rm BBN}$ and MWDS+$Y_{\rm P}^{\rm BBN}$, where the impacts of causes 1 and 2 are negligible.

\subsection{Mirror world with physical recombination and massive neutrinos}
\label{sec:MnuAndPhysRec}

\begin{figure}[!]
\centering
\includegraphics[trim={2.7cm 4.5cm 3.8cm 6.4cm},clip,width=\columnwidth]{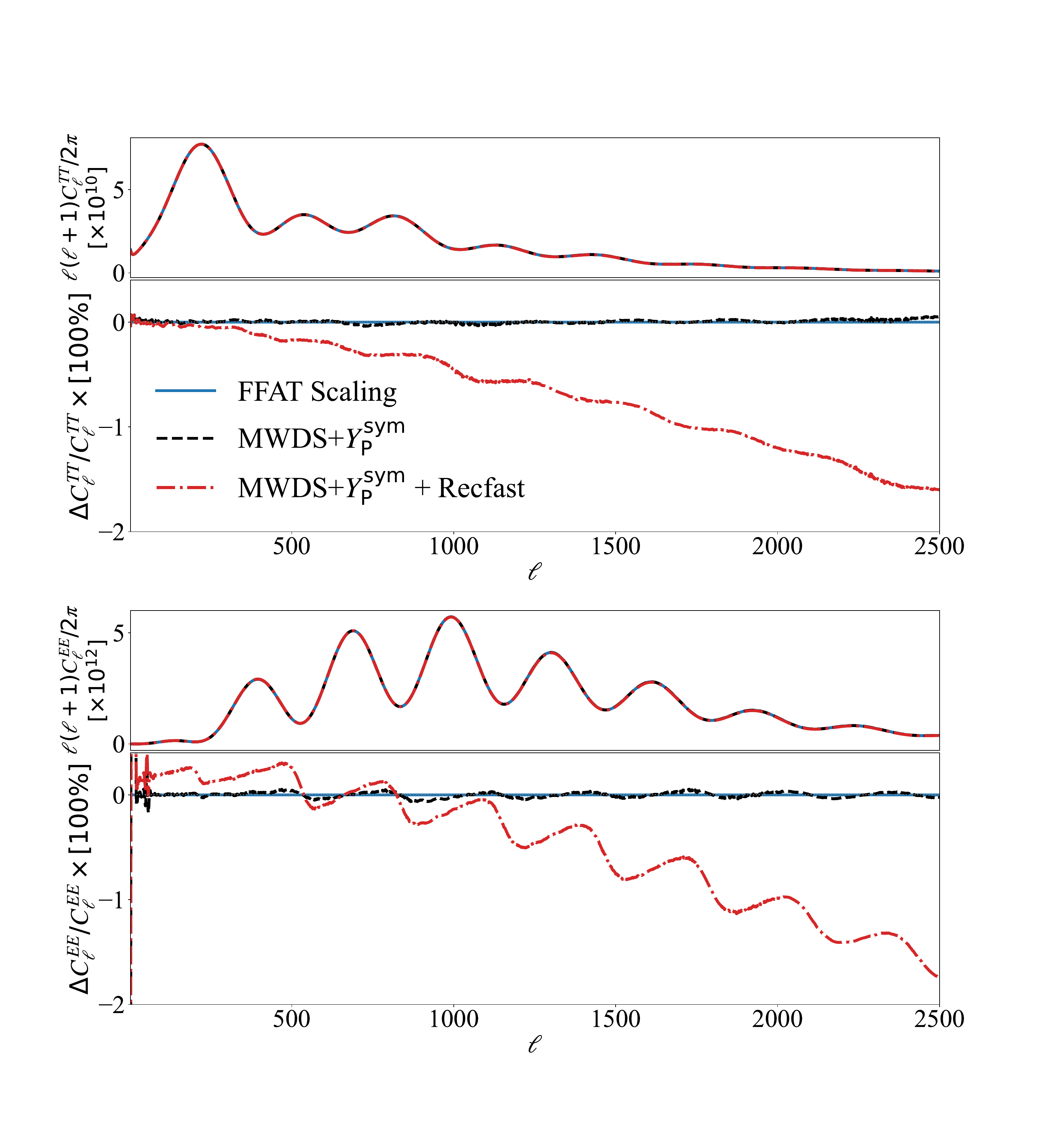}
\caption{CMB power spectra of MWDS models with and without fixing the recombination history. The CMB spectra of the FFAT scaling model are shown in blue solid lines. The MWDS + $Y_{\rm P}^{\rm Sym}$ model with fixed ionization history is shown in black dashed lines. The MWDS + $Y_{\rm P}^{\rm Sym}$ model with physical recombination process is shown in red dash-dotted lines.}
\label{fig:adm_recfast2}
\end{figure}

In the previous subsections, we have assumed massless neutrinos and a fixed ionization history for all the models. Under both assumptions, the MWDS model can exploit the scaling symmetry without violating the COBE/FIRAS constraint. However, the scaling transformation is not preserved by the physical recombination process and the presence of massive neutrinos.

The sensitivity to the expansion rate is acquired during the out-of-equilibrium recombination, with its sensitvity to the ratio of the expansion rate to the microphysical reaction rates, as emphasized by ZZ. When the expansion rate is faster, the recombination will deviate from the equilibrium state earlier. Since the photon diffusion damping of the power spectra and the CMB polarization anisotropy are generated during this out-of-equilibrium period, the physical recombination process may not preserve the dimensionless CMB power spectra. In ZZ, they emphasize the physical recombination rate as responsible for the symmetry-breaking effects seen in their CMB TT and EE power spectra. However, these effects are also due to other sources of changes to the photon scattering rate relative to the Hubble rate. Here we isolate the effects just due to changing the recombination rates relative to the Hubble rate; i.e., we isolate the effects due to cause 4.

In Fig.~\ref{fig:adm_recfast2}, we compare the CMB TT and EE spectra with a physical recombination process (red dash-dotted lines) to the one with fixed recombination history (black dashed lines) and the FFAT scaling model (blue solid lines). The deviation is only about $2\%$ up to $\ell \approx 2500$. The symmetry-breaking effect from out-of-equilibrium recombination is thus mild, although not entirely negligible. We will see in the next section that the effect, the fourth in our categorization of effects, has an impact on the posterior distribution of $H_0$ probability for the MWDS + $Y_{\rm P}$ model. 

\begin{figure}[!]
\centering
\includegraphics[trim={2.7cm 4.5cm 3.8cm 6.4cm},clip,width=\columnwidth]{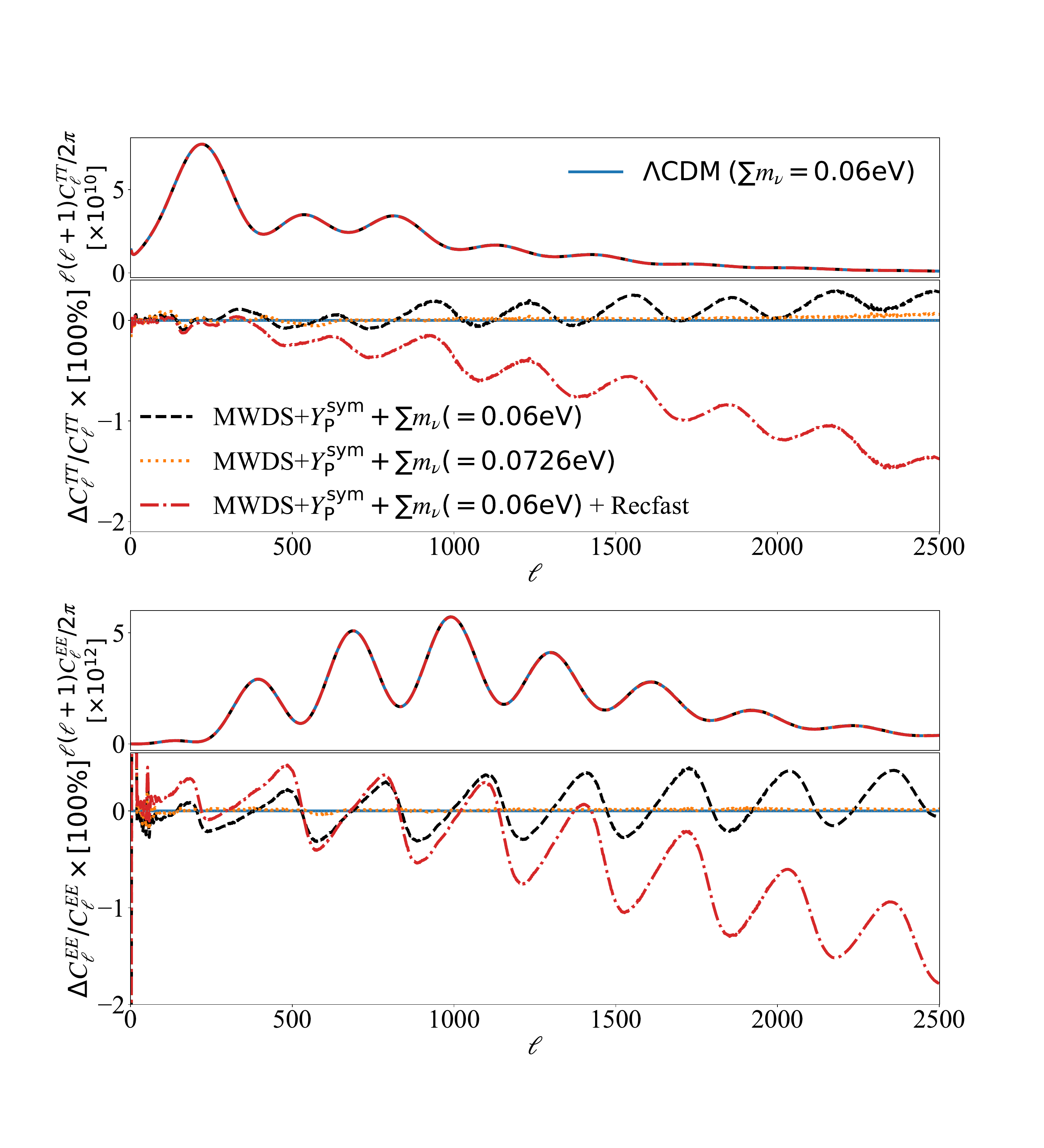}
\caption{Similar to Fig.~\ref{fig:adm_recfast2}, but showing power spectra for a \lcdm\ model with massive neutrinos and power spectra for MWDS models scaled from this \lcdm\ model, with and without fixing $x_{\rm e}(z)$ and scaling the neutrino mass.}
\label{fig:adm_massivenu}
\end{figure}
In Fig.~\ref{fig:adm_massivenu}, we show the impact of massive neutrinos on the ${\rm MWDS}+Y_{\rm P}^{\rm Sym}$ model scaled up from the best-fit \lcdm\ model with $\Sigma m_\nu = 0.06$ eV. We compare the CMB TT and EE power spectra to those of this same best-fit \lcdm\ model (blue solid lines). We see some small ($< 0.5$\%) departures from the FFAT scaling result.

These small changes emerge because the scaling of neutrino energy density is no longer uniform across redshift. Exactly how this departure from uniform scaling occurs depends on how the neutrino sector is being modeled. We have modeled it with one massive species and $N_\nu -1$ massless species. Our default scaling method leaves the massive species alone, and increases $N_\nu-1$ sufficiently to achieve, at high redshift, $\rho_\nu(z) \to \lambda^2 \rho_\nu(z)$.  When the massive neutrino becomes nonrelativistic, if we have not scaled up its mass then we have $\rho_{\nu, {\rm massive}}^{\rm non-rel}\propto \Sigma m_\nu \propto \lambda^0$, so $H(z)$ at low $z$ drops below its scaling value. 

All of the departures from FFAT scaling occur at low redshift well after recombination. Thus the impacts on the spectra have to do with late-time effects. The dominant impact  is due to a shift in the distance to last-scattering. At higher $\ell$ we also see some impact of changes to gravitational lensing. Gravitational lensing is sensitive to $H(z)$ through its impact on distance ratios, as well as on the growth of structure \cite[e.g.][]{Pan:2015bgi}.

The degree of symmetry can be improved by extending the scaling transformation to include $m_\nu \rightarrow \lambda^2 m_\nu$. This does not fully restore the symmetry, but one can see in Fig.~\ref{fig:adm_massivenu} (see line with $\Sigma m_\nu = 0.0726$ eV) that it brings us very close. This scaling is possible as long as the absolute scale of neutrino masses is unknown. Once determined through direct laboratory measurements (see e.g.~Ref.~\cite{KATRIN:2019yun}), this new absolute scale would provide a new source of FFAT scaling breaking, albeit at a very mild level. 

\section{Constraints on Light Relics} \label{sec:constraints}
In the previous section we studied power spectrum differences that arise with a best-fit \lcdm\ model when there are additional light relics. We looked at these differences for various light relic model spaces. Here we discuss the constraints on light relic energy densities, as parametrized by $N_{\rm eff}$, as well as the Hubble constant $H_0$.

For brevity, we do not cover the same range of model spaces as in the previous section. We focus on four which have progressively looser constraints on $N_{\rm eff}$:
\begin{itemize}
    \item $N_{\rm fs}$: We assume free-streaming light relics only and allow $N_{\rm fs}$ to vary. We also set the primordial helium fraction to the BBN-predicted value.
    \item $N_{\rm fs}+Y_{\rm P}$: We assume free-streaming light relics only and allow $N_{\rm fs}$ to vary. We also set the primordial helium fraction, $Y_{\rm P}$, free.
    \item $N_{\rm fs}+N_{\rm fld}+Y_{\rm P}$: We assume that the light relics consist of free-streaming and fluidlike species, and set both $N_{\rm fs}$ and $N_{\rm fld}$ independently free. We also set $Y_{\rm P}$ as a free parameter.
    \item Mirror World+$Y_{\rm P} $ (MWDS+$Y_{\rm P}$): We allow the scaling transformation factor $\lambda$ and the mirror world dark baryon fraction of total dark matter $f_{\rm ADM}$ to vary independently with a flat prior range $1.00001<\lambda<1.3$ and $f_{\rm ADM} \in [0,1]$. We also allow $Y_{\rm P}$ to vary independently. The dark photon temperature is set by $T^D_{\gamma}/T_{\gamma}=(\lambda^2-1)^{1/4}$. 
\end{itemize}

\subsection{Constraints from CMB and BAO}
\label{sec:cosntraint_cmbbao}

We first look at the constraints from CMB and BAO. The light relics $N_{\rm eff}$ includes the free-streaming species ($N_{\rm fs}$), fluidlike additional relics ($N_{\rm fld}$) and dark photons ($N_{\gamma}^D$); i.e., $N_{\rm eff} = N_{\rm fs}+N_{\rm fld}+N_{\gamma}^D$. In all four models above, we calculate the physical ionization history, $x_{\rm e}(z)$, using RECFAST \citep{Seager:1999bc, Wong:2007ym} and calculate the ionization history of the mirror world following \citep{Cyr-Racine:2013ab}. The sum of neutrino masses, $\sum {m_\nu}$, is also fixed at 0.06 eV.

To get the constraints on $N_{\rm eff}$ and $H_0$, we modified CAMB \citep{Lewis:1999camb} to solve the relevant Boltzmann equations and used CosmoMC \citep{Lewis:2002mc} to sample the parameter posterior distribution. We combine the CMB data (Planck TT+TE+EE, Lowl$_{\rm T}$, Lowl$_{\rm E}$ and lensing likelihood \citep{Planck:2019nip}) and baryon acoustic oscillation (BAO) data (6dFGS \citep{Beutler:2011}, SDSS MGS \citep{Ross:2014qpa} and BOSS DR12 \citep{BOSS:2016wmc}) to get joint constraints.

\begin{figure}[]
\includegraphics[width=\columnwidth]{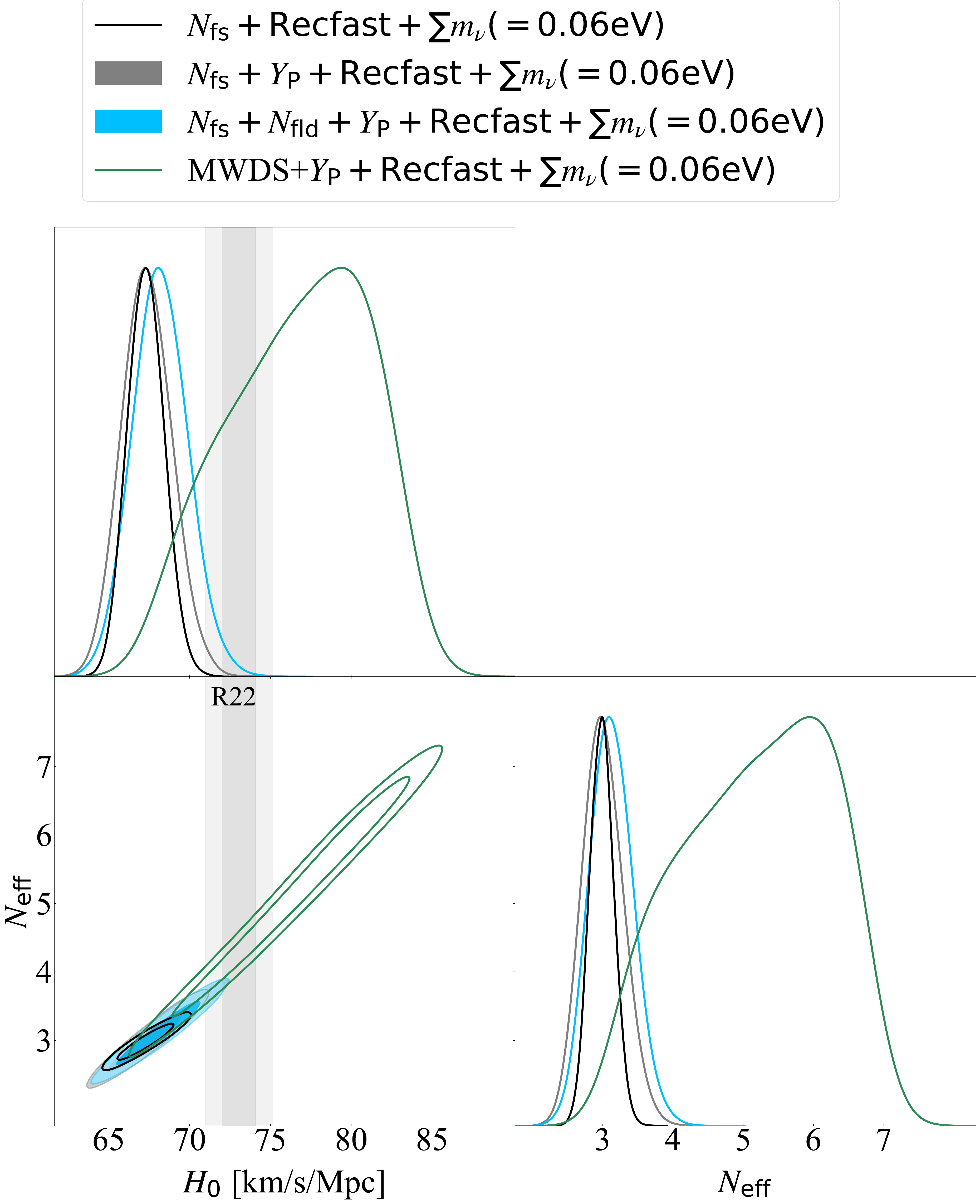}
\caption{Constraints on the effective number of neutrino species ($N_{\rm eff}$) and the Hubble constant ($H_0$). The free-streaming species, fluidlike species, and dark photons all  contribute to $N_{\rm eff}=N_{\rm fs}+N_{\rm fld}+N_{\gamma}^D$. In black are the constraints given the model with only free-streaming species and BBN-predicted $Y_{\rm P}$ [$N_{\rm fs}+{\rm Recfast}+\Sigma {m_\nu}\ (=0.06{\rm eV})$], while the constraints given the model with only free-streaming species and free $Y_{\rm P}$ are shown in gray [$N_{\rm fs}+Y_{\rm P}+{\rm Recfast}+\Sigma {m_\nu}\ (=0.06{\rm eV})$]. Results given the model with free $N_{\rm fs}$, $N_{\rm fld}$ and $Y_{\rm P}$ are shown in blue [$N_{\rm fs}+N_{\rm fld}+Y_{\rm P}+{\rm Recfast}+\Sigma {m_\nu}\ (=0.06{\rm eV})$]. The green curves show the constraints of the MWDS $+Y_{\rm P}+{\rm Recfast}+\Sigma  {m_\nu}\ (=0.06{\rm eV})$ model. In these models, the primordial helium abundance, $Y_{\rm P}$, is also a free parameter. The total mass of neutrinos, $\Sigma m_\nu$ is set to $0.06 \ {\rm eV}$ in all three models. The gray band in the left two panels shows the $H_0$ measurement from \cite{Riess:2021jrx}.}
\label{fig:chains}
\end{figure}

The constraints on $N_{\rm eff}$ and $H_0$ of the four models are shown in Fig.~\ref{fig:chains}. We see for the first three model spaces the uncertainties opening up moderately under the progression from one, two, and then three beyond-\lcdm\ parameters. The progression is expected both from a pure parameter-count perspective, and also because of the new scaling transformations each additional parameter allows. 

With each new parameter added, the allowed scaling transformations share more of the properties with MWDS + $Y_{\rm P}^{\rm Sym}$ scaling, as one can see in Table~\ref{tab:models}. The more these properties are shared, the fewer the differences between the reference model and the corresponding point\footnote{Recall that this corresponding point has the same value of $N_{\rm eff}$ as the reference model and is reached by scaling from \lcdm\ with the scaling transformation that has the highest degree of symmetry allowed in the model space.} in the model space of interest. 

In Fig.~\ref{fig:chains} we see a dramatic expansion of uncertainties as we move from those first three models to the MWDS + $Y_{\rm P}$ model. This model allows for $N_{\rm eff}$ to be increased along a direction in parameter space (the MWDS + $Y_{\rm P}$ (with physical $x_{\rm e}(z)$) scaling direction) that avoids causes 1 to 3. The difference with the result for the $N_{\rm fs} + N_{\rm fld} + Y_{\rm P}$ model, which allows for avoidance of causes 1 and 2, is due to the impact of cause 3, as discussed in the previous section. 

That there should be such a dramatic expansion of uncertainties is not at all obvious from the results we saw in the previous section. The four model spaces we consider here, ordered from most constraining to least, allow for elimination of none of our listed causes, cause 1, causes 1 and 2,  and causes 1, 2, and 3. We found in the previous section that cause 1, 2, 3, and 4 lead to differences between the \lcdm\ spectra and spectra of models with $N_{\rm eff}= 4.61$ ($\lambda = 1.1$) at the 10 to 15\% level, 5 to 6\% level, 2 to 3\% level and 1 to 2\% level respectively. Just based on these numbers alone one might expect at most a doubling of uncertainty in going from our second-to-least constraining model to our least constraining model. 

Recall though that these spectral comparisons were done along these scaling directions. There is nothing to guarantee that the maximally symmetric scaling transformation takes us from best-fit \lcdm\ to the best-fit location in the new model space. In general, there will be variations of parameters that take us off that scaling trajectory that can act to reduce the residuals displayed in the previous section. These accidental degeneracy directions (accidental as they are not associated with any known transformation symmetry) are also important to the quantitative results on display in Fig. 8. Apparently, the residuals we saw due to cause 4 are not only smaller, but also particularly amenable to reduction via these accidental degeneracy directions.

The uncertainties under the MWDS + $Y_{\rm P}$ model are not infinite. The constraining power emerges due to cause 4, and the inability of other parameter variations to completely undo its impact. Cause 4 is the recombination rates relative to the Hubble rate, which impact $x_{\rm e}(z)$ and therefore the photon scattering rate relative to the Hubble rate. This is the cause articulated by ZZ, but not distinguished by them from other, usually more dominant, causes of changes to $n_{\rm e}(z)/H(z)$.  Here we see, for the first time, the impact on parameter constraints due to cause 4 alone. They only become important in a model space in which the other three causes can be avoided. 

The mode of the one-dimensional posterior for $H_0$ given the MWDS + $Y_{\rm P}$ model is shifted to significantly higher $H_0$. The CMB data do have a slight preference for models in this space with higher $H_0$. The best-fit model has $H_0 = 79.69$ and a $\chi^2$ value that is lower by 4 from that of the best-fit \lcdm\ model. This level of improvement, given four additional free parameters, is consistent with noise fitting expectations given \lcdm\ as the true model. 

The residuals seen in Sec.~\ref{sec:MnuAndPhysRec} due to the impact of cause 4, though small, are nevertheless significant given the Planck error bars. The best-fit MWDS + $Y_{\rm P}$ model has much smaller residuals. Evidently the residuals in Fig.~\ref{fig:adm_recfast2} can be mostly compensated with the appropriate parameter adjustments. 

\subsection{Additional constraints from stellar ages and light-element abundances}
\label{sec:cosntraint_agebbn}

The MWDS + $Y_{\rm P}$ model opens up constraints on $N_{\rm eff}$ to such a large degree that it is important to consider other sources of constraints on $N_{\rm eff}$ from observables that are not protected by the scaling transformation symmetry. We briefly consider three of those here.

\begin{figure}[!h]
\includegraphics[width=\columnwidth]{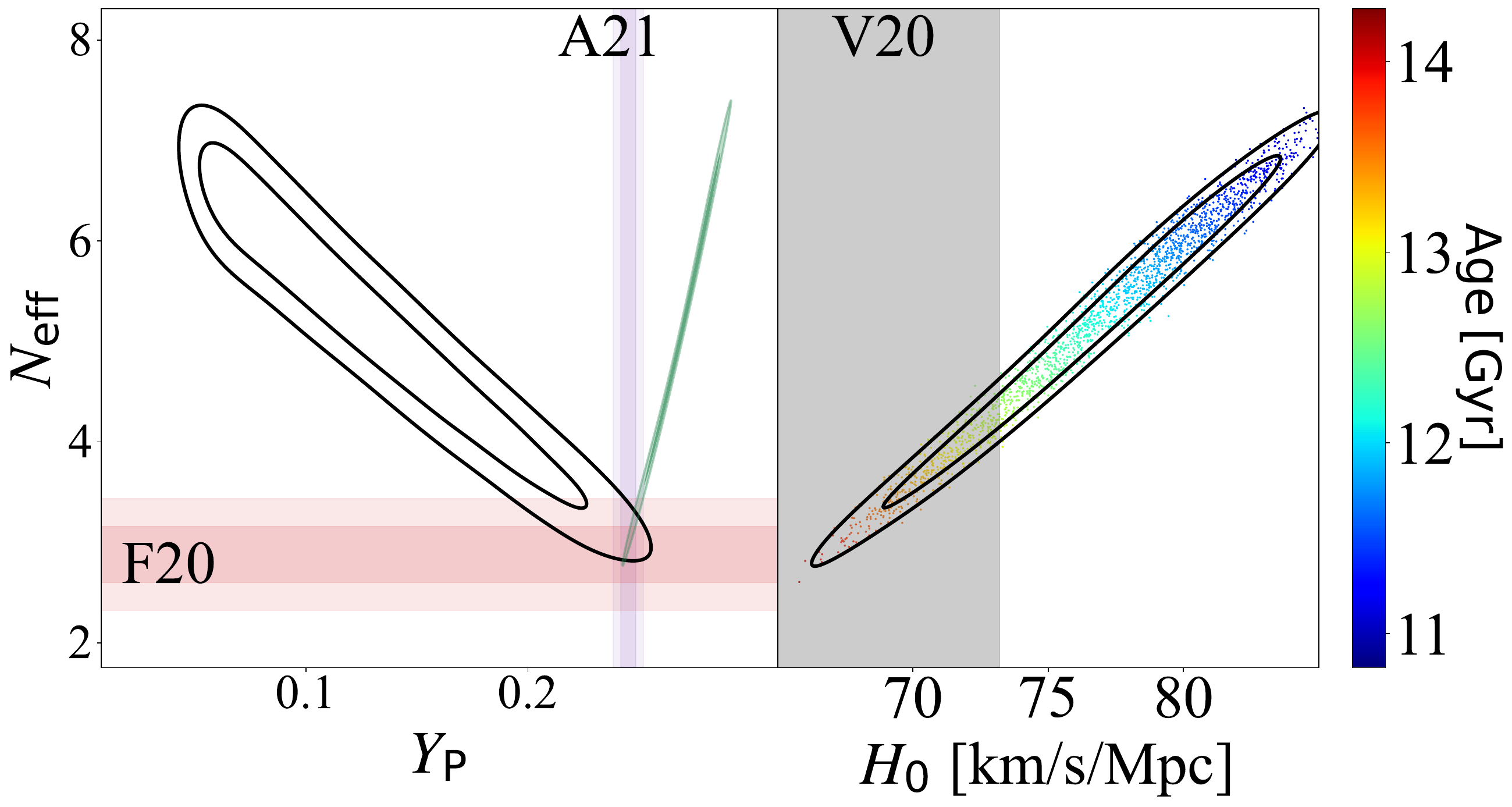}
\caption{Constraints in the $N_{\rm eff}-Y_{\rm P}$ and  $N_{\rm eff}-H_0$ planes from a variety of datasets. The black contour lines indicate the 68\% and 95\% credible regions from CMB and BAO data given the MWDS+$Y_{\rm P}$ model. In the right panel are also samples from that posterior probability distribution color coded by the age of the Universe at $z=0$. In the left panel the green contours, so tight in one direction that they appear collapsed down to a green curve, are the 68\% and 95\% credible regions for BBN-consistent helium for the same model and datasets. Also shown in the left panel are $Y_{\rm P}$ constraints from \cite{aver2021improving}, inferred from extragalactic regions of metal-poor ionized gas, and constraints on $N_{\rm eff}$ derived from inferences of helium and deuterium and BBN predictions \cite{Fields:2019pfx}. The right edge of the shaded region in the right panel is the 95\% confidence upper limit on $H_0$ from Vagnozzi et al. \cite{Vagnozzi:2021tjv} based on inferred ages of old astrophysical objects over a range of redshifts. It is cosmology model dependent, but in a way that makes it appropriate for our application here, as we describe in the text. 
}
\label{fig:age_bbn}
\end{figure}

There is a long tradition of inferring the primordial helium abundance from observations of metal-poor ionized gas. Spectral observations in the ultraviolet include both He and H transitions. These can be modeled to simultaneously determine properties of the medium, including the ratio of helium to hydrogen. The low metallicity indicates relatively small amounts of stellar processing, allowing one to extrapolate helium inferences over a range of metallicities to a primordial helium abundance (at zero metallicity with small uncertainty. Aver et al.~\cite{aver2021improving} report $Y_{\rm P} = 0.2453\pm0.0034$ (see also Refs.~\cite{Izotov:2014fga,Aver:2015iza,2018MNRAS.478.5301F,Hsyu:2020uqb}). We can see in the left panel of Fig.~\ref{fig:age_bbn} that the helium measurement, combined with the Planck measurements, place a very strong constraint on $N_{\rm eff}$ for the MWDS + $Y_{\rm P}$ model. This constraint could be circumvented with an alternative mechanism, besides lowering $Y_{\rm P}$, to scale up the photon scattering rate [CGK]. 

Yet further constraints from light element primordial abundance determinations emerge when one considers their creation in the first few minutes of the Universe. With some standard assumptions (such as zero neutrino chemical potential), one can calculate expected light element production as a function of $N_{\rm eff}$ and the baryon-to-photon ratio. For baryon-to-photon ratios consistent with Planck observations, Fields et al.~\cite[][hereafter F20]{Fields:2019pfx} infer $N_{\rm eff} = 2.88 \pm 0.28$ from measurements of primordial helium and deuterium abundances. 

These bounds could be circumvented with a violation of some of the assumptions of the standard BBN calculation. One of these assumptions is that the $N_{\rm eff}$ relevant for CMB anisotropy calculations is the same as for BBN calculations. A late reheating of the MWDS could potentially evade these constraints (see e.g.~Refs.~\cite{Aloni:2021eaq,Joseph:2022jsf,Buen-Abad:2022kgf}), although no viable scenario has yet been described. 

The scaling transformation $H(z) \rightarrow \lambda H(z)$ also sends the age of the Universe at any redshift  $t_U(z) \rightarrow t_U(z)/\lambda$. Lower limits on the age at any redshift thus place upper bounds on $\lambda$. 
Vagnozzi et al.~\cite{Vagnozzi:2021tjv} have recently considered constraints on ages from modeling of old astrophysical objects observed over a range of redshifts at $z < 8$. Their constraints are cosmology model dependent, with sensitivity to the shape of $H(z)$ at these low redshifts. However, the shape they assume is that of \lcdm , with $\Omega_{\rm m}$ assumed to be in a fairly narrow range, with the amplitude controlled by $H_0$. That is the same shape of $H(z)$ preserved by our scaling transformations. Given how tightly the data constrain the MWDS + $Y_{\rm P}$ model to the scaling transformation, it is a good approximation to simply apply their resulting constraint on $H_0$ directly, as we have in the left panel of Fig.~\ref{fig:age_bbn}.

\section{Summary and Conclusions}
\label{sec:conclusions}

In this paper we introduced a new general framework for understanding constraints on light relics from CMB observations using a set of scaling transformations. The scaling transformations all involve various rates that impact the evolution of cosmological perturbations, as well as an $n_{\rm s}$-dependent scaling of the amplitude of the primordial density perturbation power spectrum. They range from the least-comprehensive scaling that merely preserves the shape of the Hubble rate [$H(z) \rightarrow \lambda H(z)$] to the most-comprehensive case of a uniform scaling of {\em all} the relevant rates in the problem [with recombination rates effectively scaled by an artificial fixing of the ionization fraction history $x_{\rm e}(z)$]. This latter scaling leaves all dimensionless cosmological observables invariant, as we pointed out in CGK.  

The gravitational free-fall rates, $\sqrt{G\rho_i(z)}$ for each component $i$ , the photon (Thomson) scattering rate $\sigma_{\rm T}n_{\rm e}(z)$, and hydrogen recombination rates, are all the rates that enter the evolution equations for \lcdm\ and the extensions we consider here.
The FFAT scaling transformation (implicitly understood to be at fixed $x_{\rm e}(z)$) is a uniform scaling of all these rates.  

The FFAT scaling transformation is tightly constrained by the FIRAS determination of the mean photon density today and therefore its associated free-fall rate today. The MWDS + $Y_{\rm P}$ model, introduced in CGK, can evade this constraint and allows for a scaling transformation that provides very nearly the same degree of symmetry as FFAT scaling. The MWDS includes dark baryons and dark photons. This model admits the MWDS + $Y_{\rm P}^{\rm Sym}$ scaling transformation that we showed to be essentially equivalent to FFAT scaling regarding its impact on CMB spectra. We used this scaling transformation to identify a reference model, useful for understanding constraints on light relics in the various model spaces we consider.

We identified four causes of observational consequences that can lead to constraints on light relics. These causes are all associated with a lack of scaling of some rate, or a nonuniform scaling of rates. They are:
 1) changes to the photon scattering rate relative to the Hubble rate, 2) changes to the fraction of radiation density that is freely streaming, 3) changes to the fraction of nonrelativistic matter that is pressure supported, and 4) changes to recombination rates relative to the Hubble rate. The ordering here is from most impactful causes to least impactful. 
The first two are well-studied in the literature \cite{Hu:1996,Bashinsky:2003tk,Hou:2011ec,Baumann:2015rya}, the third we articulated here for the first time, and the fourth was originally proposed in Ref.~\cite{Zahn_2003}.

We used the scaling transformations to create, in Sec.~\ref{sec:anal_models}, informative comparisons of spectra from different model spaces. We found there that cause 1, 2, 3, and 4 lead to differences between the \lcdm\ spectra and spectra of models with $N_{\rm eff}= 4.61$ ($\lambda = 1.1$) at the 10 to 15\% level, 5 to 6\% level, 2 to 3\% level and 1 to 2\% level respectively. 

To investigate how cause 3 leads to observable consequences we 
compared a model with increased $N_{\rm eff}$, reached by a scaling transformation that preserves the rate ratios of causes 1 and 2 (the Mix+$Y_{\rm P}^{\rm Sym}$ scaling transformation), to its corresponding reference model. We found that this model, after horizon crossing, has reduced gravitational potential decay, relative to the reference model, due to all of its dark matter lacking pressure support, while in the reference model a fraction of the dark matter has pressure support. This difference in potential evolution changes the CMB power spectra from the \lcdm\ best-fit one mainly through its impact on the monopole source of the CMB spectra by shifting the equilibrium position of the monopole oscillation and driving the photon perturbation itself to a larger amplitude during the subhorizon evolution period. The dark baryons in the MWDS model introduce pressure support to a non-zero fraction of the dark matter, making evolution of the gravitational potential appropriate for preserving CMB anisotropy and polarization observables under the FFAT scaling transformation.

Cause 4 was articulated first by Ref.~\cite{Zahn_2003}, but its observable consequences are only now clear. This is the only one of the four causes that cannot be eliminated by the scaling transformations possible in the MWDS + $Y_{\rm P}$ model. We saw its impact on spectral differences in Sec.~\ref{sec:anal_models} and, more indirectly, on the constraints on $N_{\rm eff}$ and $H_0$ in Sec.~\ref{sec:constraints}. 

In principle we could list an additional cause that can be important in model spaces with nonzero neutrino mass. We found though that a fixed neutrino mass, of 0.06 eV, only mildly breaks the symmetry, at the fractions of a percent level for $\lambda = 1.1$. This mild symmetry breaking is due to small departures in $H(z)$ at late time from its scaling trajectory.

Because CMB and BAO data allow for fairly high values of $N_{\rm eff}$ and $H_0$ in the MWDS + $Y_{\rm P}$ model we reported, in Sec.~\ref{sec:cosntraint_agebbn}, on constraints on this model space from other observables, that greatly restrict this freedom. Primordial helium abundance measurements, comparison of light element abundance measurements and BBN predictions, and inferences of the ages of the oldest astrophysical objects, all lead to significant constraints on $N_{\rm eff}$. 

Our analysis has proceeded from identification of the key rates in the problem that make sensitivity to $H(z)$ from dimensionless observables, such as CMB spectra, possible. These are gravitational free-fall rates, the photon scattering rate, and recombination rates. Although we have restricted our analysis to extensions with additional light relics, we note that in {\em any} model space, the ability to predict $H_0$ given CMB spectra also depends fundamentally on known rates that control out-of-equilibrium processes. Similar analyses thus may help to bring analytic understanding to constraints from a broader set of cosmological models.

\acknowledgments
 LK and FG were supported in part by DOE Office of Science award DE-SC0009999.   F.-Y.~C.-R is supported by the National Science Foundation (NSF) under grant AST-2008696. LK acknowledges the hospitality of TTK at RWTH University where some of this work was completed.

\appendix
\section{The Impact of the Source Terms and CMB Lensing on CMB Power Spectra} \label{apd:cls}

In this appendix, we discuss the contributions of the effects induced by the gravitational potential change to the overall difference between the Mix+$Y_{\rm P}^{\rm Sym}$ model and the FFAT scaling and MWDS + $Y_{\rm P}^{\rm Sym}$ models. We will discuss the contributions of the monopole term, Doppler term and integrated Sachs-Wolfe effect sourcing the CMB spectrum first. Then we show the impact from CMB lensing.

\begin{figure}[!]
\includegraphics[trim={0cm 0.5cm 1.5cm 1.5cm},clip,width=\columnwidth]{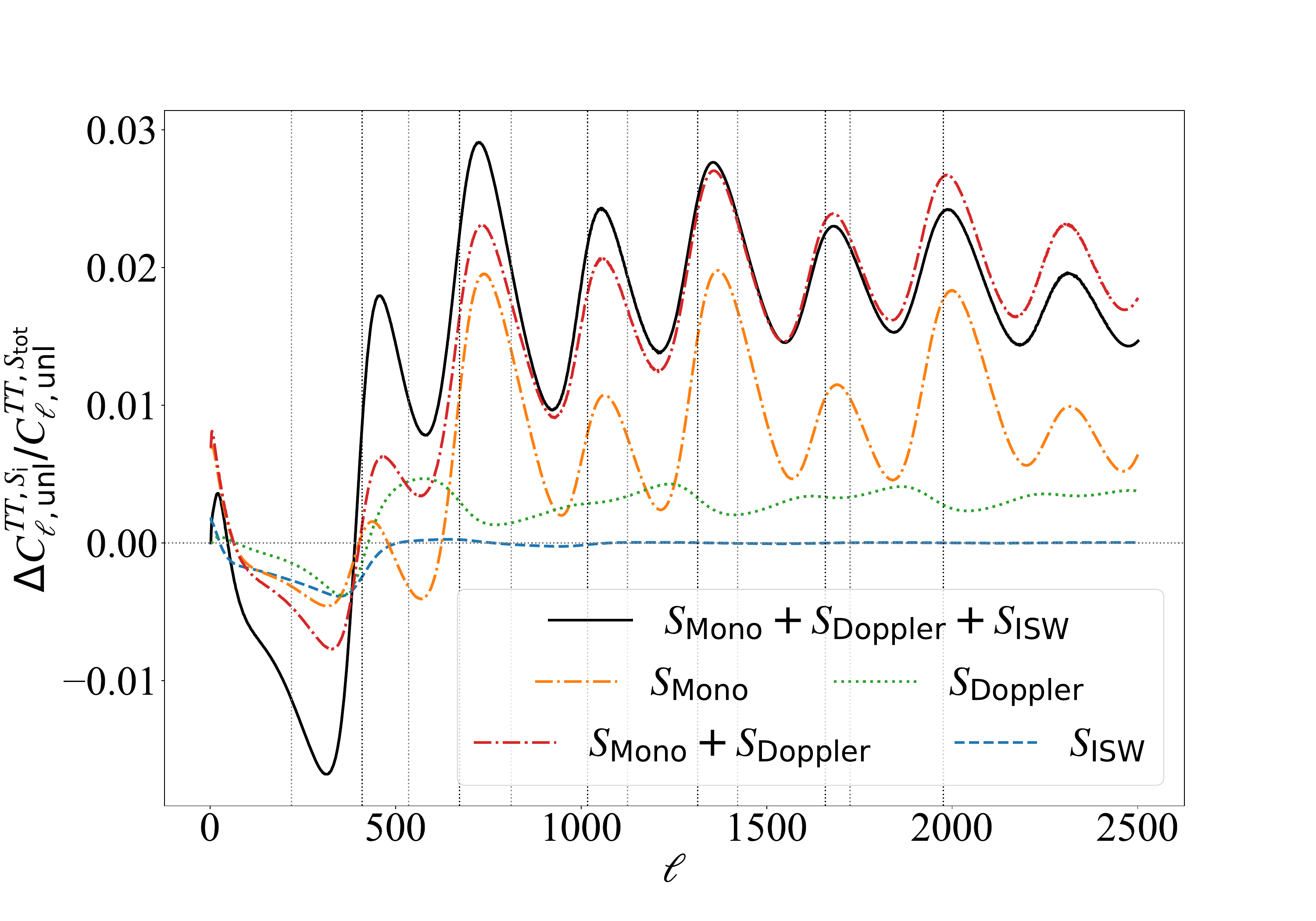}
\caption{Fractional contributions of different source terms to the difference in the unlensed CMB power spectrum between the Mix+$Y_{\rm P}^{\rm Sym}$ model and the FFAT scaling model. The gray vertical lines show the peak locations in CMB power spectrum and the black vertical lines show the trough locations.}
\label{fig:source_analysis}
\end{figure}

To start with, we can regroup the CMB temperature anisotropy source term into the following form:
\begin{align}
    S(k, \eta) \approx& S_{\rm Mono} + S_{\rm Doppler} + S_{\rm ISW} \\
    \approx& g(\eta)[\Delta_{T0}(k, \eta)+\psi(k, \eta)]+\frac{1}{k^2}\frac{d}{d \eta}[\theta_{\rm b}(k, \eta) g(\eta)] \nonumber\\ 
    &+e^{-\tau}[\dot{\psi}(k, \eta)+\dot{\phi}(k, \eta)]\nonumber.
\end{align}
The three main contributions to the CMB power spectrum are the monopole ($S_{\rm Mono}$), the Doppler term ($S_{\rm Doppler}$) and the ISW effect ($S_{\rm ISW}$). In Fig.~\ref{fig:source_analysis}, we show the fractional differences of the source terms contributing to the overall difference in the unlensed CMB power spectra between the Mix+$Y_{\rm P}^{\rm Sym}$ model and the FFAT scaling model. 

\begin{figure}[!]
\includegraphics[trim={0.5cm 0cm 0cm 0cm},clip,width=\columnwidth]{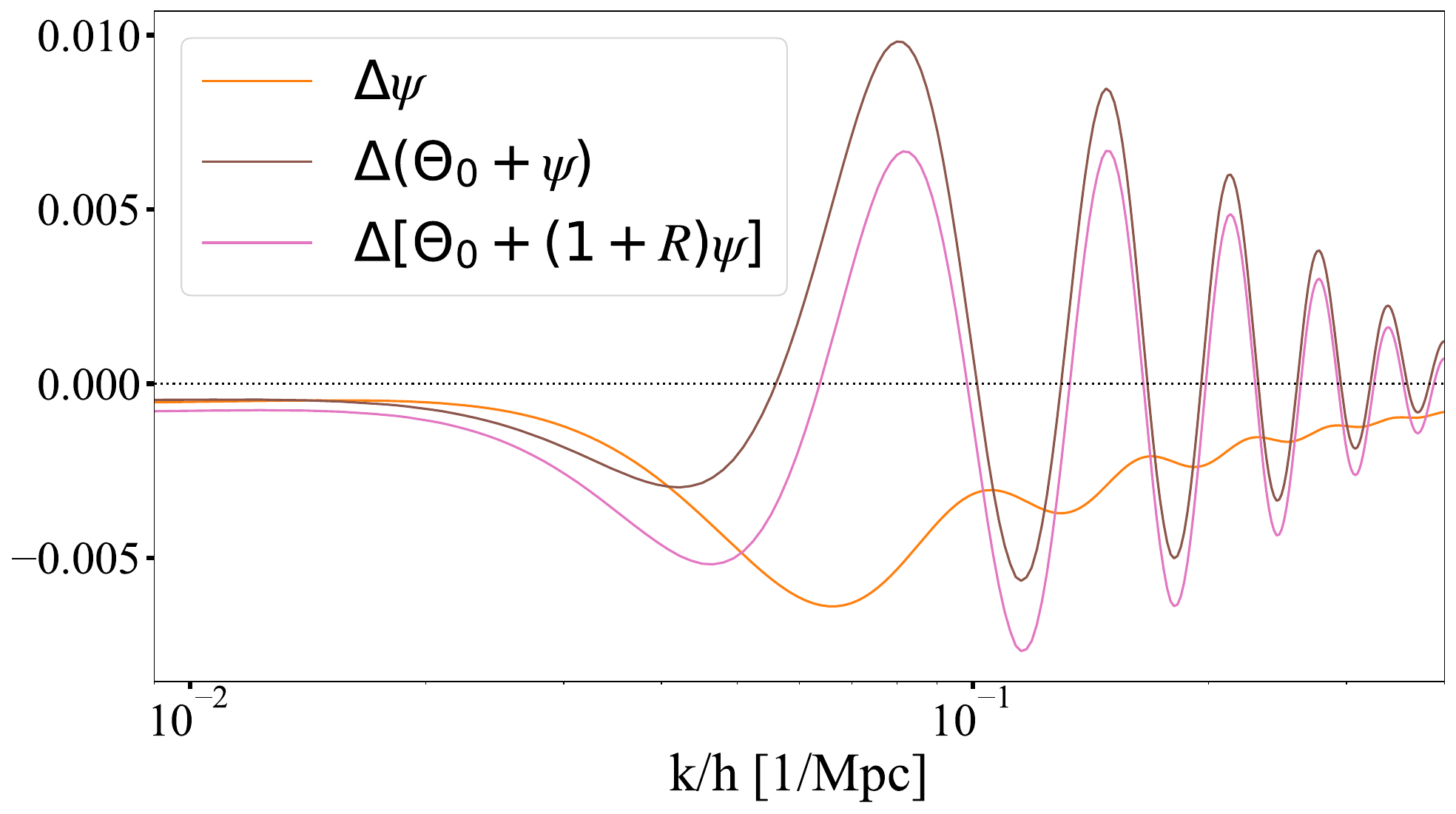}
\caption{Residuals of Mix+$Y_{\rm P}^{\rm Sym}$ model with respect to MWDS+$Y_{\rm P}^{\rm Sym}$ model at the time of last scattering as a function of $k/h$. The baryon-to-photon density ratio factor, $R$, at last scattering is 0.61. }
\label{fig:mono_residual}
\end{figure}

At scales $\ell \gtrsim 600$, the leading difference comes from the monopole power ($S_{\rm Mono}$, in orange), while the contribution from the Doppler term ($S_{\rm Doppler}$, in green) is subdominant. Compared to the power from monopole source alone, the power of monopole plus Doppler source ($S_{\rm Mono}+S_{\rm Doppler}$, in red) have larger boost at monopole power troughs than monopole power peaks. This is due to the fact that the Doppler term is not in phase with the monopole oscillation. The ISW power (in blue) are almost the same at this range between the two models. The difference of the power spectra of the complete sources (in black) to the $S_{\rm Mono}+S_{\rm Doppler}$ only spectra is due to cross terms with the ISW source in the power spectra.

The contributions of the monopole term are twofold. The additional cold dark matter without pressure support in Mix+$Y_{\rm P}^{\rm Sym}$ induces a deeper gravitational potential, shifting the zero-point equilibrium of oscillator $\Theta_0+\psi$ further away from zero. On small scales corresponding to the third and higher peaks, where the subhorizon evolution is long enough, the photon perturbation, $\Theta_0$, itself is driven to a larger amplitude due to reduced potential decay. In Fig.~\ref{fig:mono_residual}, we show the residuals of $\psi$, $\Theta_0+\psi$ and $\Theta_0+[1+R(z_*)]\psi$ at last scattering as a function of $k/h$. The zero-point equilibrium position of $\Theta_0+\psi$ is $-R\psi$, and the baryon-to-photon density ratio factor, $R$ is 0.61 at last scattering in the base \lcdm\ model. We see that $\Delta \left[\Theta_0+(1+R)\psi\right]$ oscillates around zero. The alternating high-low peaks of the monopole power seen in Fig.~\ref{fig:source_analysis} are due to the change to the zero-point equilibrium position of $\Theta_0+\psi$ oscillator. The increase to the overall amplitude of $\Theta_0+\psi$ in Mix+$Y_{\rm P}^{\rm Sym}$ model leads to the overall positive fractional changes to the monopole spectrum at high-$\ell$ range.

On large scales, however, the contributions from the monopole, Doppler and ISW term are comparable.

\begin{figure}[!]
\vspace*{3mm}
\includegraphics[width=\columnwidth]{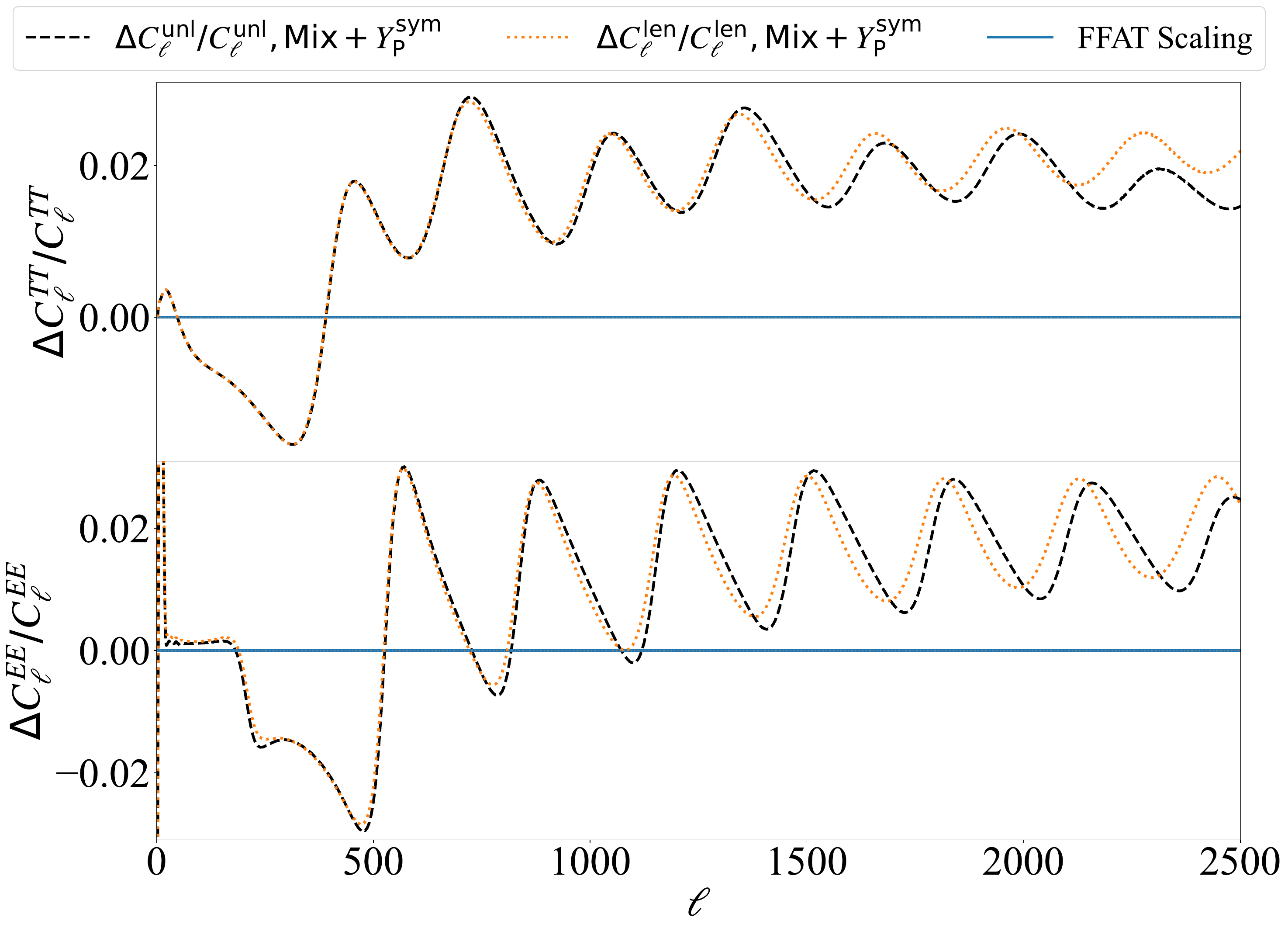}
\caption{Fractional changes of the lensed and unlensed power spectra from Mix+$Y_{\rm P}^{\rm Sym}$ model to the FFAT scaling model.}
\label{fig:lensing_analysis}
\end{figure}

Next, we investigate the impact of CMB lensing on the difference of the CMB power spectra between the Mix+$Y_{\rm P}^{\rm Sym}$ model and the FFAT scaling model. The effect of CMB lensing on the damping tail is to ``transport" the power on large scales to small scales \citep{Lewis:2006fu}. In Fig.\ref{fig:lensing_analysis}, we compare the fractional difference in the lensed (orange lines) and unlensed (black lines) power spectra of the Mix+$Y_{\rm P}^{\rm Sym}$ model to the FFAT scaling model. The fractional differences of the lensed and the unlensed power spectra are almost of the same overall amplitude. The CMB lensing is not the main contribution to the difference in the power spectra of the two models.

\bibliography{main}
\end{document}